\pdfoutput=1
\documentclass[superscriptaddress,showpacs,aps,prd,twocolumn]{revtex4-1}

\usepackage[colorlinks=true, pdfstartview=FitV, linkcolor=red, citecolor=blue,
urlcolor=blue]{hyperref}
\usepackage[dvipdfmx]{graphicx}
\usepackage{amsmath,amssymb}
\usepackage{color}

\newcommand{\EM}{\gamma_{\text{E}}}

\newcommand{\Lag}{\mathcal{L}}

\newcommand{\bL}{\boldsymbol{L}}
\newcommand{\bOmega}{\boldsymbol{\Omega}}
\newcommand{\bv}{\boldsymbol{v}}
\newcommand{\bB}{\boldsymbol{B}}
\newcommand{\bE}{\boldsymbol{E}}
\newcommand{\bF}{\boldsymbol{F}}
\newcommand{\bnabla}{\boldsymbol{\nabla}}
\newcommand{\Veff}{V_\text{eff}}
\newcommand{\mc}{m_{\rm current}}
\newcommand{\mdyn}{m_\text{dyn}}
\newcommand{\Lpt}{\Lambda_\text{PT}}

\usepackage[normalem]{ulem}

\newcommand\rout{\bgroup \color{red} \ULdepth=-.5ex \ULset}
\newcommand\bout{\bgroup \color{blue} \ULdepth=-.5ex \ULset}

\begin{document}

\title{Analogy between rotation and density for Dirac fermions
       in a magnetic field}

\author{Hao-Lei Chen}
\affiliation{Physics Department and Center for Particle Physics and
             Field Theory, Fudan University, Shanghai 200433, China}
\author{Kenji Fukushima}
\affiliation{Department of Physics, The University of Tokyo,
             Tokyo 113-0033, Japan}
\author{Xu-Guang Huang}
\affiliation{Physics Department and Center for Particle Physics and
             Field Theory, Fudan University, Shanghai 200433, China}
\author{Kazuya Mameda}
\affiliation{Department of Physics, The University of Tokyo,
             Tokyo 113-0033, Japan}
\begin{abstract}
 We analyse the energy spectra of Dirac fermions in the presence of
 rotation and magnetic field.  We find that the Landau degeneracy is
 resolved by rotation.  A drastic change in the energy dispersion
 relation leads to the ``rotational magnetic inhibition" that is a
 novel phenomenon analogous to the inverse magnetic catalysis in a
 magnetic system at finite chemical potential.
\end{abstract}

\maketitle

\section{Introduction}

In many quantum theories an external magnetic field is a useful probe
for various intriguing phenomena.  The most important concept to
understand the magnetic dynamics appears from the Landau quantization.
For a strong enough magnetic field only the lowest Landau level (LLL)
dominates the dynamics.  Such a situation with a gigantic magnetic
field could be realized in the Early
Universe~\cite{Cheng:1994yr,*Baym:1995fk,Grasso:2000wj} and in central
cores of neutron stars (or magnetars) where
$eB\sim 10^{15}\,\text{G}$~\cite{Duncan:1992hi}.  Also, an extreme
environment with a strong magnetic field could be generated in
relativistic heavy-ion collision experiments where we may have
$eB\sim m_\pi^2 \sim 10^{18}\,\text{G}$
\cite{Skokov:2009qp,Voronyuk:2011jd,Deng:2012pc}.
Investigating quantum chromodynamics (QCD) in a strong magnetic field
is, therefore, of increasing importance for not only theoretical
interest but also experimental application~\cite{Brambilla:2014jmp}.
In particular the response of quark matter to the magnetic field
involves $\mathcal{P}$- and $\mathcal{CP}$-odd processes through
quantum anomaly, which is still under active
studies~\cite{Kharzeev:2007tn,*Kharzeev:2007jp,%
*Fukushima:2008xe,Son:2004tq,Metlitski:2005pr,Son:2009tf,
Kharzeev:2010gd} (see also
Refs.~\cite{Fukushima:2012vr,*Kharzeev:2013ffa,*Huang:2015oca,%
*Kharzeev:2015znc} as related reviews).

One of the most essential and established changes of quark matter
driven by strong magnetic fields is the inevitable breaking of chiral
symmetry, which is called the magnetic
catalysis~\cite{Klimenko:1991he,*Klimenko:1992ch,Gusynin:1994re,%
*Gusynin:1995nb,Shovkovy:2012zn,*Miransky:2015ava}.  We can confirm the magnetic catalysis in many
theoretical examples which include:  the Nambu--Jona-Lasinio (NJL)
model~\cite{Gusynin:1994re,*Gusynin:1995nb,Ebert:1999ht,Inagaki:2003yi},
the quark-meson model~\cite{Fraga:2008qn,Mizher:2010zb,Andersen:2011ip,Ferrari:2012yw}, the MIT
bag model~\cite{Fraga:2012fs}, the lattice QCD
simulation~\cite{Bali:2012zg}, the holographic
model~\cite{Johnson:2008vna} (see also
Refs.~\cite{Shovkovy:2012zn,*Miransky:2015ava} for reviews and the
references therein).  The idea of charged particles acquiring a
dynamical mass due to the magnetic field is applicable also to
condensed matter systems such as the
graphene~\cite{Gorbar:2007xh,*Gorbar:2008hu}, the Weyl
semimetals~\cite{Roy:2014mia}, and bosonic
systems~\cite{Ayala:2012dk,Feng:2014bpa}.

The magnetic catalysis could be affected by other controlling
parameters even though the magnetic field is strong.  For instance, in
a dense system at large chemical potential, the magnetic field would
not enhance but suppress the chiral condensate, which is called the
inverse magnetic catalysis~\cite{Ebert:1999ht,Preis:2010cq}. Inclusion
of neutral meson fluctuations could lead to an infrared singularity that
disfavors the chiral condensate, which is called the magnetic
inhibition~\cite{Fukushima:2012kc}.  In this way it would be useful to
consider magnetic systems under competing conditions for drawing
further nontrivial consequences from the physics of the magnetic
catalysis
\cite{Bali:2011qj,Fukushima:2012xw,Preis:2012fh,Bruckmann:2013oba}.

In this work we will pay attention to the competition between
rotation and the magnetic catalysis.  It is well known that the effect
of rotation or angular momentum is quite analogous to that of the magnetic
field.  Especially for nonrelativistic systems in a trapping
potential one can show that the system exhibits the Landau-type
quantization in response to
rotation~\cite{Wilkin:2000zz,*Cooper:2001zz,*Schweikhard:2004zz,%
Mameda:2015ria}.  This analogy has motivated people to study anomalous
quantum phenomena induced by rotation instead of magnetic field, that
is, the quantum Hall effect induced by
rotation~\cite{viefers2008quantum}, the quantum vortex with rotating
Bose-Einstein condensate~\cite{tsubota2013novel}, the chiral
vortical effect~\cite{Son:2009tf}, and the chiral magnetic effect in cold atoms~\cite{Huang:2015mga}.

We would emphasize another interesting (and less known) analogy
between rotation and density.  For nonrelativistic theories this
analogy is readily understood from the fact that the Hamiltonian in a
rotating frame is shifted as
$\hat{H} \to \hat{H}-\hat{\bL}\cdot\bOmega$ (with the angular velocity
vector $\bOmega$) and this latter term may be regarded as an effective
chemical potential.  One might thus expect that the similarity between
rotation and density should hold for relativistic theories.  However,
the similarity in the relativistic case is, if any, not as trivial
because one should treat relativistic rotation as a deformation of
spacetime geometry.  That means, to study rotational effects on
relativistic systems, it is necessary to analyse the quantum field
theory in curved spacetime~\cite{parker2009quantum}. (Also from the 
viewpoint of the Poincar\'{e} algebra, rotating relativistic fluid can 
be discussed~\cite{Becattini:2007zn,Becattini:2007nd,*Becattini:2009wh,%
*Becattini:2013fla}.)  Although QCD in
curved spacetime is not yet a mature research subject and not much
about modified QCD vacuum structure is known, it has been argued
that the gravitational background fields should significantly
influence the QCD vacuum
properties~\cite{Elizalde:1993kb,*Inagaki:1993ya,*Inagaki:1997kz,
*Gorbar:1999wa,*Huang:2006fk,
Schutzhold:2002pr,*Urban:2009yg,Flachi:2011sx,
*Flachi:2013iia,*Flachi:2014jra,Benic:2015qha}
(see also Refs.~\cite{Yamamoto:2013zwa,Yamamoto:2014vda,Benic:2016kdk} for
quantum lattice simulations).  It is, therefore, an intriguing
question that whether rotation could be given an interpretation as an
effective chemical potential even for relativistic theories.  If so,
rotation should yield a modification on the QCD vacuum and,
particularly in the presence of strong magnetic field, we may
anticipate an effect analogous to the inverse magnetic catalysis in
which the role played by the chemical potential is replaced with
rotation.  We would call the phenomenon of reduced chiral condensate by a
combination of rotation and magnetic field the ``rotational magnetic
inhibition'' in short.

In this paper, we investigate the Dirac equation with both rotation
and magnetic field and apply the resulting energy dispersion relation
to a fermionic effective model.  The solution of the Dirac equation
indicates that the modified Landau levels with rotation have
nondegenerate spectrum with angular momentum dependence.  We adopt the
NJL model and impose both the magnetic field and rotation to find
chiral restoration that is driven by increasing magnetic field
especially at strong coupling.  Finally we will discuss possible
physical implications of our results to several experimental setups.

\section{Dirac equation in a rotating frame}

In curved spacetime generally the Dirac equation with electromagnetic
fields can be written as
\begin{equation}\label{eq:generalDeq}
  \bigl[ i\gamma^\mu (D_\mu + \Gamma_\mu ) - m \bigr]\psi = 0
\end{equation}
with the covariant derivative $D_\mu \equiv \partial_\mu + ieA_\mu$ and $e>0$ being the charge of the Dirac fermion.  As usual, the affine
connection $\Gamma_\mu$ is defined in terms of the metric $g_{\mu\nu}$
or the spin connection $\omega_{\mu ij}$ and the vierbein $e^\mu_i$ as
\begin{equation}
\begin{split}
  & \Gamma_\mu = -\frac{i}{4}\omega_{\mu ij}\sigma^{ij} \;,\\
  & \omega_{\mu ij} = g_{\alpha\beta}e^\alpha_i
    (\partial_\mu e^\beta_j + \Gamma^\beta_{\mu\nu}e^\mu_j ) \;,\\
  & \sigma^{ij} = \frac{i}{2}[\gamma^i,\gamma^j] \;.
\end{split}
\end{equation}
The Greek and the Latin letters denote the indices in coordinate and
tangent space, respectively.

We can implement rotation by specifying the metric characterized by
the angular velocity vector, ${\boldsymbol \Omega} = \Omega \hat z$,
and the metric then takes the following form:
\begin{equation}\label{eq:metric}
  g_{\mu\nu} = \begin{pmatrix}
    1-(x^2+y^2)\Omega^2 & y\Omega & -x\Omega & 0 \\
    y\Omega & -1 & 0 & 0 \\
    -x\Omega & 0 & -1 & 0 \\
    0 & 0 & 0 & -1 \\ \end{pmatrix} \;.
\end{equation}
In the following calculation we adopt
\begin{equation}
  e^t_0 = e^x_1 = e^y_2 = e^z_3 = 1 \;,\quad
  e^x_0 = y\Omega \;, \quad e^y_0 = -x\Omega \;,
\end{equation}
and the other components are zero, which gives the
metric~\eqref{eq:metric}.  We shall choose the symmetric gauge in the
inertial frame and use the vector potential, $A_i=(0,By/2,-Bx/2,0)$,
which results in $\bB = B\hat z$ with $B$ a constant.  We can then give an explicit form
of the Dirac equation under rotation and the magnetic field, that is,
\begin{align}
  & \Big[ i\gamma^0(\partial_t - x\Omega\partial_y
     + y\Omega\partial_x - i\Omega\sigma^{12})
     + i\gamma^1(\partial_x + ieBy/2) \notag\\
  & \qquad + i\gamma^2(\partial_y - ieBx/2)
     + i\gamma^3\partial_z - m \Big] \psi = 0 \;.
\label{eq:Deq}
\end{align}
We can solve this differential equation to obtain the wave-function as
explained in Appendix~\ref{app:solution}.  For our purpose to study the
vacuum structure, we do not need the wave function but only the energy
spectrum is sufficient.

It is easy to deduce the eigen-energies of Eq.~\eqref{eq:Deq} at
finite $\Omega$ from the $\Omega = 0$ case.  In this case the problem
is reduced to solving the ordinary Dirac equation in an external
magnetic field.  It is a well-known fact that charged spin-$s$
particles have the energy dispersion relation in $\bB=B\hat z$
($eB>0$) as
\begin{equation}\label{eq:DROmega0}
  E^2 = p_z^2 + (2n+1-2s_z)eB + m^2 \\
\end{equation}
with non-negative integer $n$.  Compared with that without rotation,
the Dirac equation~\eqref{eq:Deq} with rotation has additional pieces
of
\begin{equation}\label{eq:Omegaj}
  -i(x\Omega\partial_y + y\Omega\partial_x)
  + \Omega\sigma^{12} = \Omega (\hat L_z + \hat S_z ) \;.
\end{equation}
We denote the eigenvalues for $\hat L_z$ and $\hat S_z$ as $\ell$ and
$s_z$, respectively.  We can regard $E + \Omega (\ell + s_z )$ as the
energy eigenvalue in the inertial frame.  In this way we can reach the
expression of the energy dispersion relations from Eq.~\eqref{eq:Deq}
given by
\begin{equation}\label{eq:DR}
  \Big[E+\Omega (\ell+s_z)\Big]^2 = p_z^2 + (2n+1-2s_z)eB + m^2 \;.
\end{equation}
In what follows we discuss some features of Dirac fermions in a
rotating frame.
\vspace{1em}

(I) First, we make a comment on the Lorentz force in a rotating frame.
The gauge fields are transformed in a rotating frame into the
following form:
\begin{equation}
  A_\mu = A_i e_\mu^i= (-B\Omega r^2/2,By/2,-Bx/2,0) \;,
\end{equation}
which leads to an electric field;
$\bE = -\bnabla A_0 = B\Omega(x,y,0)$.  Hence, na\"{i}vely, one may
want to identify this $\bE$ as the Lorentz force:
\begin{equation}
  \bF = e\bv \times \bB = eB\Omega (x,y,0) \;,
\end{equation}
where $\bv=\Omega(-y,x,0)$ is the velocity vector at $(x,y,0)$ caused
by rotation.  However, $A_0=-B\Omega r^2/2$ does not appear in
Eq.~\eqref{eq:Deq} because the gamma matrix
$\gamma^t = \gamma^i e^t_i$ cancels it out.  Therefore, rotation does
not induce any electromagnetic effect.  This is an important point
that ensures our later discussion on the similarity between rotation
and finite density for relativistic theories.
\vspace{0.5em}

(II) Let us take a closer look at the comparison of
Eqs.~\eqref{eq:DROmega0} and \eqref{eq:DR}.  Without rotation,
Eq.~\eqref{eq:DROmega0} expresses the ordinary Landau quantization in
which the motion on the $xy$-plane is characterized by $n$ only
instead of $(p_x,p_y)$.  Each Landau level has degeneracy associated
with some quantum number;  when the area of the $xy$-plane is $S$, the
degeneracy factor for each Landau level is gauge independent and given
by
\begin{equation}\label{eq:degeneracy}
  N = \biggl\lfloor \frac{eBS}{2\pi} \biggr\rfloor \;.
\end{equation}
In the cylindrical coordinates, for example, the degenerate quantum
number is the canonical angular momentum $\ell$.
Thus, $\ell$ should take $N$ different integers.  In addition, a
condition, $\ell\geq -n$ for the $n$th Landau level arises from
normalizability of the wave-function.  It follows that the possible
range of $\ell$ should be
\begin{equation}\label{eq:lrange}
  -n \leq \ell \leq N - n
\end{equation}
for the $n$th Landau level.  We give a detailed derivation for this
in Appendix~\ref{app:degeneracy}.  (One might think that $\ell$ should run
up to $N-n-1$ but we implicitly assume $N\gg 1$ with sufficiently
strong magnetic field.)

The angular momentum in Eq.~\eqref{eq:DR} also runs from $-n$ to $N-n$
(see Appendix~\ref{app:solution} for details).  This means that we have
nondegenerate spectrum depending on $\ell$ in Eq.~\eqref{eq:DR}.  We
should replace the phase space integration with double sum with
respect to $n$ and $\ell$.  For spin-$1/2$ fermions with up spin, the phase space sum reads:
\begin{equation}\label{eq:phase-space}
\begin{split}
  & \text{Magnetic field:} \int \frac{dp_xdp_y}{(2\pi)^2} \to
    \frac{eB}{2\pi}\sum_{n=0}^\infty\;, \\
  & \text{Magnetic field + Rotation:} \int \frac{dp_xdp_y}{(2\pi)^2} \to
    \frac{1}{S}\sum_{n=0}^\infty \sum_{\ell=-n}^{N-n} \;,
\end{split}
\end{equation}
where $N$ is the Landau degeneracy factor given in
Eq.~\eqref{eq:degeneracy}.  This modified phase space sum is needed
for the evaluation of the thermodynamic potential.
\vspace{0.5em}

(III) The analogy between rotation and density is clear from
Eq.~\eqref{eq:DR}.  The dispersion relation~\eqref{eq:DR} behaves as
if the Dirac fermion were put at finite density with a chemical
potential $\mu_j = \Omega j \equiv \Omega(\ell+s_z)$. Note that due 
to this similarity, the Dirac fermions under rotation also suffer from
the sign problem for Monte Carlo simulations~\cite{Yamamoto:2013zwa}. 
In this paper, motivated by such a similarity, we study a relativistic 
many-body system with rotation and magnetic field.  In the next section 
we will show that rotation may supersede the magnetic catalysis.
\vspace{0.5em}

(IV) Finally, we address the necessary condition for the system size.
For quantization in harmonic oscillators, the system size should be
large enough as compared to typical scales of the problem.  In order
to discuss the Landau quantization in the cylindrical system with area
$S=\pi R^2$, hence, the radius $R$ should be larger than the magnetic
length, $\ell_B = 1/\sqrt{eB}$ (see Appendix~\ref{app:solution}).  In the
rest frame we have no problem taking such a large cylinder toward the
thermodynamic limit, i.e.\ $R\to\infty$.  Once rotation is involved, however, the system with infinitely large radius is
not well-defined because the causality might be violated at the edges
of the cylinder where
$R\Omega\geq 1$~\cite{Davies:1996ks,Duffy:2002ss}.  Therefore, our
treatment in this paper is legitimate if $R$ is not too small to
justify the quantization, and not too large to maintain the causality.
That is, the following condition should be imposed:
\begin{equation}\label{eq:condition}
  1/\sqrt{eB} \ll R \leq 1/\Omega\;.
\end{equation}
We note that $N\gg 1$ follows from the above condition.

\section{NJL model with rotation and magnetic field}

We investigate the dynamical breaking of chiral symmetry in the
presence of rotation and magnetic field using the NJL
model~\cite{Nambu:1961tp}, which is defined in curved spacetime by
\begin{align}
 S &= \int d^4x \sqrt{-\det (g_{\mu\nu})}\, \Lag(\bar\psi,\psi), \notag\\
 \Lag &= \bar{\psi}\big[i\gamma^\mu(D_\mu+\Gamma_\mu)-\mc\big]\psi\notag\\
  &\quad\quad\quad\quad\quad\quad\quad\quad  + \frac{G}{2} \big[(\bar\psi\psi)^2+(\bar\psi\gamma_5\psi)^2 \big] \;.
\label{eq:NJL}
\end{align}
Here, $\det (g_{\mu\nu}) = -1$ for the metric~\eqref{eq:metric} and $G$
denotes the coupling constant.  In the usual way we can introduce
auxiliary fields and utilize the mean-field approximation to obtain
the effective thermodynamic potential:
\begin{equation}\label{eq:Veff}
\begin{split}
  \Veff(m) = &\frac{(m-\mc)^2}{2G} - \frac{T}{S}\sum_{q=\pm}
    \int_{-\infty}^\infty\frac{dp_z}{2\pi} \sum_{n,\ell,s_z} \\
  & \times \biggl\{\frac{\beta(\varepsilon+q\Omega j)}{2}
    + \ln\bigl[1+e^{-\beta (\varepsilon+q\Omega j)}\bigr]\biggr\} \;,
\end{split}
\end{equation}
where $\varepsilon$ is the energy dispersion relation without
rotation, i.e.\ $\varepsilon \equiv \sqrt{p_z^2+(2n+1-2s_z)eB + m^2}$.
The effective potential~\eqref{eq:Veff} is the same as that at finite
density once $\Omega j$ is identified as a constant chemical potential
$\mu$~\cite{Ebert:1999ht}.  We should note that we implicitly assume a
spatially homogeneous chiral condensate so that the dynamical mass in
the energy dispersion relation takes an ordinary form.  At zero
temperature, particularly, we can decompose Eq.~\eqref{eq:Veff} into
the pure-magnetic and rotational contributions as
\begin{equation}\label{eq:VeffzeroT}
  \Veff(m) = \frac{(m-\mc)^2}{2G} + V_0 + V_\Omega \;,
\end{equation}
where
\begin{equation}\label{eq:V0}
  V_0 = - \frac{eB}{2\pi} \sum_{n=0}^\infty \alpha_n
   \int_{-\infty}^\infty \frac{dp_z}{2\pi} \sqrt{p_z^2+m_n^2} \;,
\end{equation}
and
\begin{equation}\label{eq:VOmega}
\begin{split}
  V_\Omega = & -\frac{1}{S}\sum_{n=0}^\infty \alpha_n
    \sum_{\ell=-n}^{N-n} \theta(\Omega |j| -m_n) \\
  &\qquad\times \int_{-k_{nj}}^{k_{nj}} \frac{dp_z}{2\pi}
   \Bigl[ \Omega|j| - \sqrt{p_z^2 + m_n^2} \Bigr]
\end{split}
\end{equation}
with $j=\ell+1/2$ hereafter and $\Omega>0$.  Here, we introduced the 
following notations: $\alpha_n = 2 -\delta_{n0}$,
$m_n^2 = 2neB + m^2$, and $k_{nj}\equiv\sqrt{(\Omega j)^2-m_n^2}$.


Now we must specify the ultraviolet regularization scheme needed for
Eq.~\eqref{eq:VeffzeroT}.  The NJL model is not renormarizable, and so
the regularization scheme is a part of the model definition.  Without
electromagnetic background fields, a sharp cutoff is one of the most
conventional choices in the NJL model studies.  Because the sharp
cutoff is incompatible with gauge invariance, a smooth cutoff such as
the proper-time method~\cite{Schwinger:1951nm} and the Pauli-Villars
regularization would be more suitable for problems with
electromagnetic background fields.  For instance, in the derivations
of the magnetic catalysis in
Refs.~\cite{Klimenko:1991he,Gusynin:1994re}, the proper-time method
was used to regularize the pure-magnetic potential~\eqref{eq:V0}.

As long as our main concern is about the magnetic catalysis, it is
known that a na\"{i}ve cutoff scheme would yield qualitatively correct
results, as was also mentioned in Ref.~\cite{Gorbar:2011ya} (see also
Ref.~\cite{Fukushima:2012xw} in which a na\"{i}ve cutoff was adopted
with the functional renormalization group equation).  We note
that the ultraviolet divergent structure is the same regardless of
whether the system has rotation or not, apart from the $\ell$-sum, so
that we can use a na\"{i}ve cutoff to find results that physically
make sense.  Then, to make the $p_z$-integral and the $n$-sum
restricted in a region around $p_z^2 + 2neB \lesssim\Lambda^2$, we
introduce a smoothed cutoff function as~\cite{Gorbar:2011ya}
\begin{equation}\label{eq:cutoff-func}
  f(p_z,n;\Lambda) = \frac{\sinh(\Lambda/\delta\Lambda)}
    {\cosh[\tilde{\varepsilon}(p_z,n)\Lambda/\delta\Lambda]
     + \cosh(\Lambda/\delta\Lambda)}
\end{equation}
with $\tilde{\varepsilon}\Lambda\equiv\sqrt{p_z^2+2neB}$. 
We note that in this function the smoothness to exclude 
artifacts is tuned by a parameter $\delta\Lambda$. 
Actually, in the $\delta\Lambda/
\Lambda \to 0$ limit $f(p_z,n;\Lambda)$ is reduced to
the step function $\theta(1-\tilde\varepsilon)=\theta
(\Lambda^2-p_z^2-2neB)$. 
By changing $\delta\Lambda$ of such a simple 
function~\eqref{eq:cutoff-func}, we can systematically analyze 
whether our results are robust and not affected by cutoff artifact. 

For the rest of this work we will focus on $\mc=T=0$.  The gap equation is
the condition to minimize Eq.~\eqref{eq:VeffzeroT} with Eqs.~\eqref{eq:V0}
and \eqref{eq:VOmega}.  Then, we can write the gap equation down as
follows:
\begin{equation}\label{eq:gapeq}
  \frac{m}{G}  = \frac{m}{\pi}\big(F_0 - F_\Omega\big)
\end{equation}
with the pure-magnetic term given by
\begin{equation}\label{eq:Fvac}
  F_0 \equiv \frac{eB}{2\pi} \sum_{n=0}^\infty\alpha_n
   \int_0^\infty \frac{dp_z\, f(p_z,n;\Lambda)}{\sqrt{p_z^2+m_n^2}} \;,
\end{equation}
and the rotational term given by
\begin{equation}
\begin{split}
  F_\Omega \equiv & \frac{1}{S}\sum_{n=0}^\infty \alpha_n
    \sum_{\ell=-n}^{N-n} \theta(\Omega |j| -m_n) \int_0^{k_{nj}}
    \frac{dp_z\, f(p_z,n;\Lambda)}{\sqrt{p_z^2+m_n^2}} \;.
\end{split}
\label{eq:FOmega0}
\end{equation}
This expression of $F_\Omega$ is slightly complicated for the 
evaluation.
If $k_{nj}$ is  negligibly small compared with $\Lambda$ and $\sqrt{eB}$,
however, the rotational contribution $F_\Omega$ can be approximated
with a simpler regularization by $f(0,n;\Lambda)$, which significantly
simplifies the analytical treatment.  Fortunately, this is the case
for our analysis with a  small $G$ (see Sec.~\ref{subsec:weak}). In
this approximation we  can perform the $p_z$-integration in
Eq.~\eqref{eq:FOmega0} analytically to reach:
\begin{equation}
 \begin{split}
  F_\Omega &\simeq \frac{1}{S} \sum_{n=0}^\infty \alpha_n
    \sum_{\ell=-n}^{N-n}\theta(\Omega|j| -m_n)\,f(0,n;\Lambda) \\
  &\qquad\qquad\qquad \times
  \ln\biggl(\frac{\Omega|j|+\sqrt{(\Omega j)^2 -m_n^2}}{m_n}\biggr) \;.
 \end{split}
 \label{eq:FOmega1}
\end{equation}

We note that we can immediately have the expression for finite-density
systems by replacing $F_\Omega$ in Eq.~\eqref{eq:gapeq} with
\begin{equation}\label{eq:Fmu}
  F_\mu = \frac{eB}{2\pi}\sum_{n=0}^\infty\alpha_n\,
    \theta(|\mu| -m_n) 
  \ln\biggl( \frac{|\mu| + \sqrt{\mu^2 - m_n^2}}{m_n}\biggr) \;,
\end{equation}
which encompasses the mechanism for the inverse magnetic catalysis.
It should be mentioned that we would not demand an ultraviolet
regularization thanks to the step function,
$\theta(|\mu|-m_n)$ in $F_\mu$.  From this, at the same time, we can
understand that $F_\Omega$ is not really sensitive to the
regularization scheme if $S$ is large enough.

We should here refer to involved studies by Becattini $\it et\ al$. In 
Refs.~\cite{Becattini:2007nd}, the quantum relativistic fermion system 
in rotating frame has been investigated in the aim of establishing a 
general thermodynamic and hydrodynamic framework to describe the system. 
Hence the chiral symmetry breaking has not been discussed. Also they have
not considered the magnetic field which plays an important role to 
obviously show the analogy between rotation and density.

\section{Numerical Results and Discussions}

In this section we analyse the magnetic response of the dynamical
mass in rotating frames in the following two cases: (A) $G<G_c$
and (B) $G>G_c$, where $G_c$ is the critical coupling for the onset of
the chiral condensate in the vacuum (i.e.\ $\Omega = \sqrt{eB} = 0$)
\begin{equation}
  G_c=19.65/\Lambda^2\;,
\label{eq:Gc}
\end{equation}
which we numerically determined from Eqs.~\eqref{eq:gapeq} using our
present regulator \eqref{eq:cutoff-func} with
\begin{equation}
  \delta \Lambda = 0.05 \Lambda\;.
\label{eq:dLambda}
\end{equation}
We have numerically verified that a different $\delta\Lambda$ would
change the results quantitatively but the qualitative features are the
same.	For both (A) and (B) we choose the following 
parameters:
\begin{align}
  & eB=(0.1-0.2)\Lambda^2\;,\notag\\
  & S=10^6\pi/\Lambda^2 \quad (\text{i.e.\ } R=10^3/\Lambda)\;.
\label{eq:parameters}
\end{align}
We see that Eq.~\eqref{eq:condition} is satisfied with the above
parameter choice and the treatment of the Landau quantization is
justified.

\subsection{Dynamical mass at weak coupling ($G<G_c$)}
\label{subsec:weak}

\begin{figure}
  \begin{center}
    \includegraphics[width=1\columnwidth]{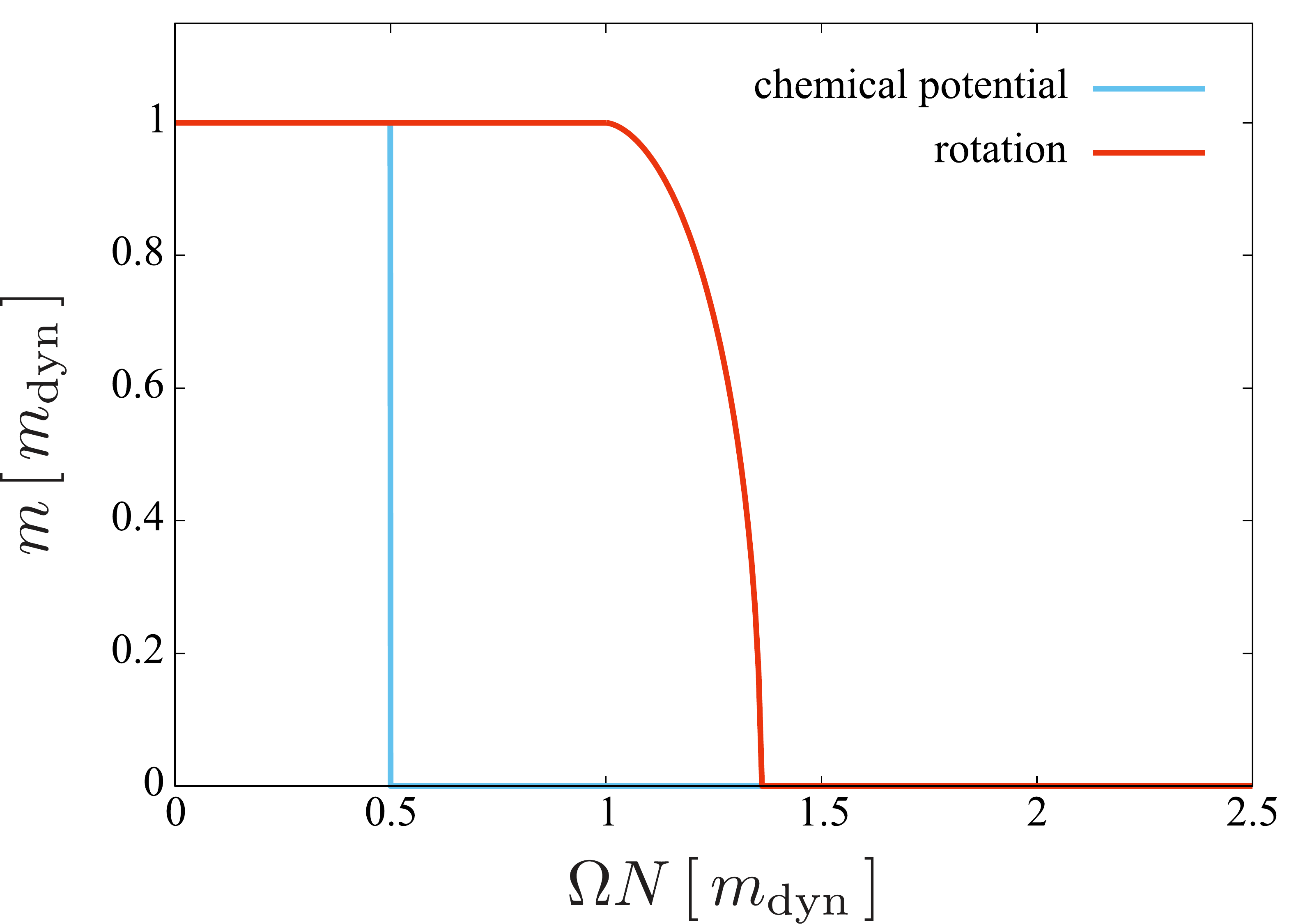}
    \caption{Dynamical mass with $eB = 0.2\Lambda^2$ obtained from the
    gap equation~\eqref{eq:gapeq} with rotation (red line) and with
    chemical potential $\mu=\mu_N$ (blue line).  The model parameters
    are chosen as in Eq.~\eqref{eq:parameters}.}
    \label{fig:rotchemi}
  \end{center}
\end{figure}

Let us discuss our numerical results with the following coupling:
\begin{equation}
 G=0.622 G_c\;.
\label{eq:weakG}
\end{equation}
We define a unit of the
dynamical mass as
\begin{equation}
  \mdyn = 1.25 \times 10^{-2} \Lambda\;,
\end{equation}
which is the solution of the gap equation with $eB = 0.2\Lambda^2$
and $\Omega = 0$.  We show
our numerical results in dimensionless unit in terms of
$\mdyn$.  In Fig.~\ref{fig:rotchemi} we make a plot for the
dynamical mass (red line) as a function of the angular velocity by
solving Eq.~\eqref{eq:gapeq} with rotation.  The horizontal axis is
given by an effective ``chemical potential'':
\begin{equation}
  \mu_N \equiv \Omega N \;.
\end{equation}
In view of Eqs.~\eqref{eq:FOmega1} and \eqref{eq:Fmu} this $\Omega N$
is the maximum counterpart of $\mu$.

\begin{figure}
  \begin{center}
    \includegraphics[width=1\columnwidth]{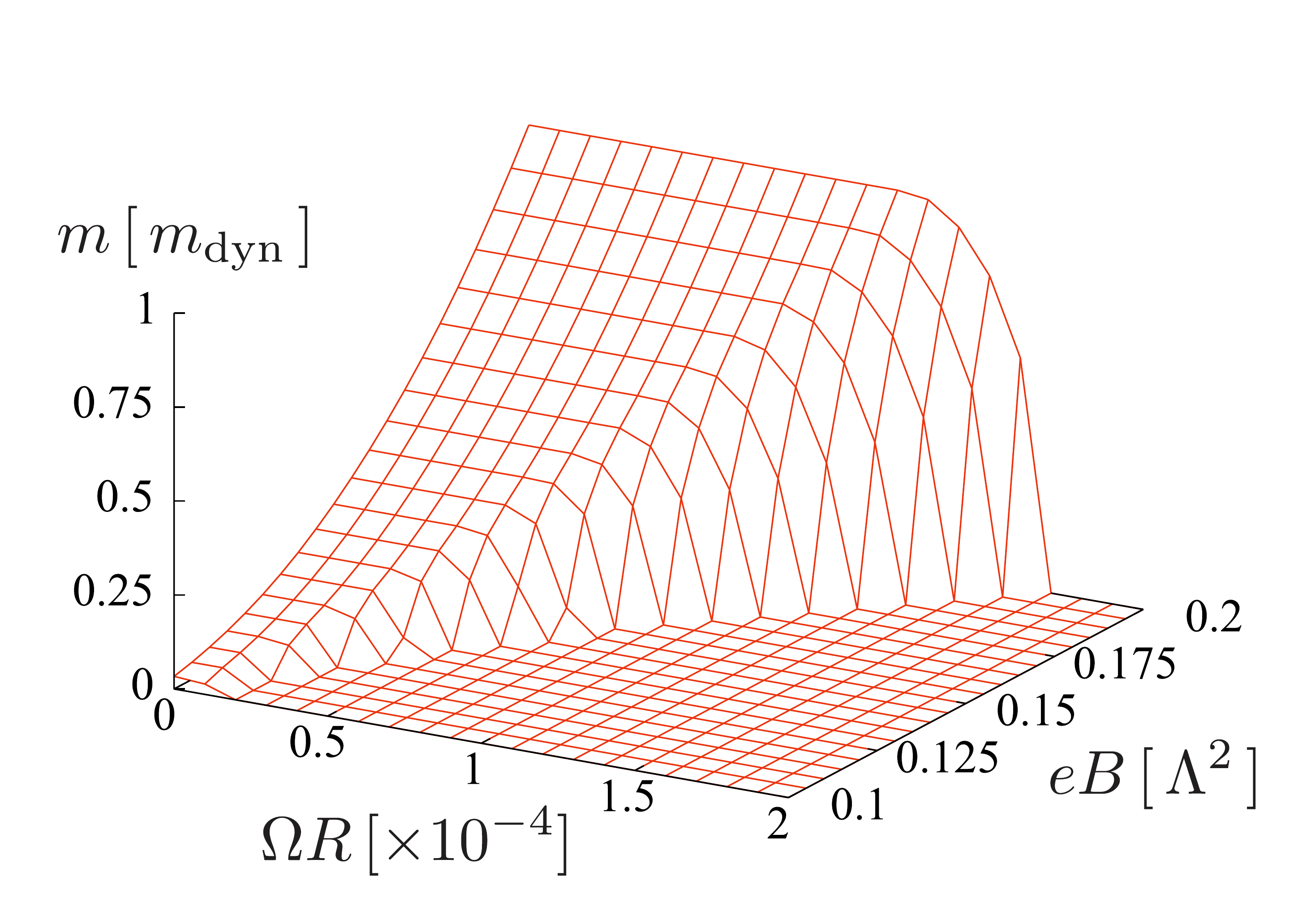}
    \caption{3D plot for the dynamical mass as a function of $\Omega$
    and $eB$ at weak coupling.  For small $\Omega$ the dynamical mass
    grows exponentially with $1/eB$ (i.e.\ the magnetic catalysis).
    The critical $\Omega_c$ is also amplified exponentially as $1/eB$
    decreases (see also Fig.~\ref{fig:Omegac}).}
    \label{fig:3dweak}
  \end{center}
\end{figure}

\begin{figure}
  \begin{center}
    \includegraphics[width=1\columnwidth]{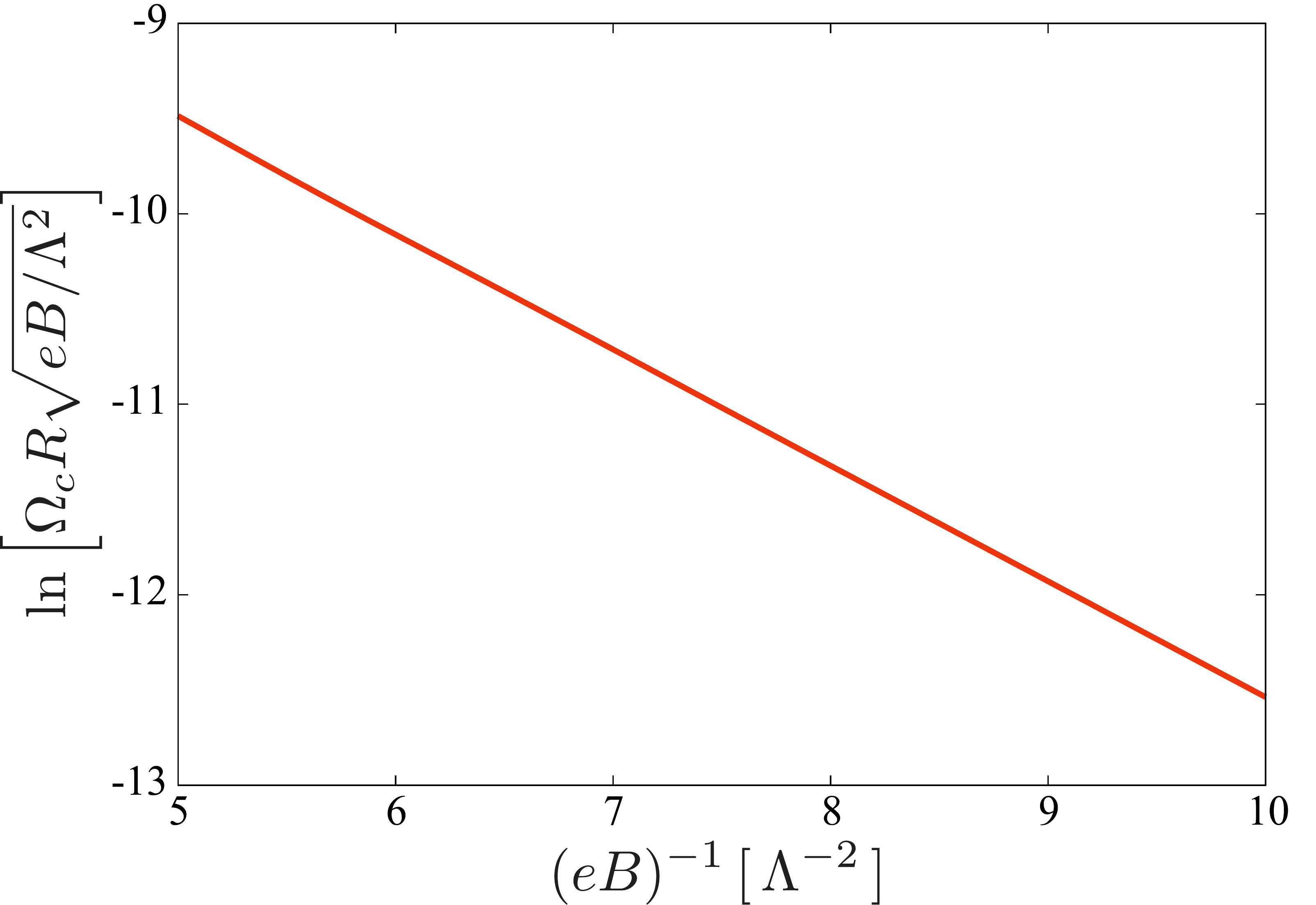}
    \caption{$eB$-dependence of $\Omega_c$ for
    $0.1\leq eB/\Lambda^2 \leq 0.2$.  The linearity of
    $\ln\Omega_c$ vs $1/eB$ confirms the validity of the functional
    form of $\Omega_c$ given by Eq.~\eqref{eq:Omegac-numerical}.}
    \label{fig:Omegac}
  \end{center}
\end{figure}

To pursue the analogy between rotation and density quantitatively, we
draw another (blue) line by solving the gap equation with $F_\Omega$
replaced with $F_\mu(\mu=\mu_N)$.  Figure~\ref{fig:3dweak} is a 3D
plot for the solution of Eq.~\eqref{eq:gapeq} as a function of
$\Omega$ and $eB$.  We can observe that there is a threshold for the
dynamical mass with increasing $\Omega$, above which $m=0$ and chiral
symmetry is restored.  This location of the critical $\Omega_c$
changes with $eB$, and we make Fig.~\ref{fig:Omegac} to show this
$eB$-dependence of $\Omega_c$.
Here are some remarks on these numerical results.
\vspace{1em}

(I) When the angular velocity exceeds $\Omega \simeq \mdyn/N$,
the rotational effects become visible, but the damping of the
dynamical mass starts slowly (see the red line in
Fig.~\ref{fig:rotchemi}).  This behavior is different from the mass
suppression induced by finite chemical potential (see the blue line in
Fig.~\ref{fig:rotchemi}).  Such a difference stems from the
$\ell$-dependence of each mode.

Let us count the number of modes that are relevant for the suppression
of the dynamical mass.  For simplicity, we concentrate only on the LLL
with $n=0$.  In the present parameter choice, the LLL approximation
should work well for $F_\Omega$ and $F_\mu$.  (The following argument
should be also applicable even when higher Landau levels are not 
negligible.)  Because of the step
function in $F_\Omega$, only the modes with $\ell>m/\Omega-1/2$ give
finite contributions.
Indeed, the red line in Fig.~\ref{fig:rotchemi} starts decreasing at
$N\Omega \simeq \mdyn$, which corresponds to the threshold that the
highest angular momentum modes in $F_\Omega$ contributes
nonvanishingly.  In contrast, the step function in $F_\mu$ given in
Eq.~\eqref{eq:Fmu} indicates that all $N$ modes simultaneously start
contributing for $\mu> m$, while for $\mu< m$ nothing happens.
\vspace{0.5em}

(II) Another way to investigate the difference between the red and the
blue lines in Fig.~\ref{fig:3dweak} is to approximate the $\ell$-sum.
Suppose that $\Omega$ is small so that we can treat $\Omega j$ as a
continuous variable.  Also we assume a sufficiently large integer
$N$.  Then, we can approximate the $\ell$-sum in $F_\Omega$ by an
integration as
\begin{equation}
  \begin{split}
   & \sum_{\ell=-n}^{N-n}\ln\left(\frac{\Omega |j| +
   \sqrt{(\Omega j)^2 - m_n^2}}{m_n}\right) \theta(\Omega|j|-m_n) \\
   & \simeq  \frac{1}{\Omega}\int_{0}^{\mu_N}d\mu\ln\left(
   \frac{\mu+\sqrt{\mu^2 - m_n^2}}{m_n}\right) \theta(\mu-m_n)\;.
  \end{split}
\label{eq:integral}
\end{equation}
For our parameter choice $N\sim \mathcal{O}(10^4)$ is large enough and the above approximation is justified.
Then the rotational contribution to the gap equation \eqref{eq:gapeq}
is reduced to
\begin{equation}
  F_\Omega = F_\mu(\mu = \mu_N) - \frac{eB}{2\pi}\sum_{n=0}^\infty
  \alpha_n\sqrt{1-\frac{m_n^2}{\mu_N^2}}\,\theta(\mu_N -m_n)\;.
\label{eq:FOmega2}
\end{equation}
It is obvious that a density-like effect induced by rotation is
certainly contained in the first term $F_\mu$.  The second is a negative
term that makes a difference from the finite-density case.  This extra
term plays a role to weaken chiral restoration by rotation as compared
to that by high density.  Therefore, the suppression of the dynamical
mass in the rotating frame occurs more gradually than that with the
finite chemical potential.  Moreover, Eq.~\eqref{eq:FOmega2} implies
$F_\Omega < F_\mu$ for a fixed $\mu_N$, and thus, chiral restoration
by rotation would need larger $\mu_N$ than that by finite density (see
Fig.~\ref{fig:rotchemi}).
\vspace{0.5em}

(III) For $\mc=T=0$ and large $eB$ we can analytically investigate the
$eB$-dependence of $\Omega_c$.  In our analysis we adopted the
na\"{i}ve cutoff regularization with Eq.~\eqref{eq:cutoff-func}, but
the regularization scheme should be irrelevant for a large system with
$S \gg 1/eB$.  If we utilized the proper time regularization for
$F_0$, the gap equation with rotation and strong magnetic field would
be~\cite{Schwinger:1951nm}
\begin{equation}
 \begin{split}
  &\frac{4\pi^2}{G} = \Lpt^2 -m^2 \biggl[\ln\biggl(
   \frac{\Lpt^2}{2eB}\biggr) - \gamma_{\text{E}}\biggr] \\
  &\qquad + eB \Bigg[ \ln \biggl(\frac{m^2}{4\pi eB}\biggr)
   + 2\ln\Gamma\biggl(\frac{m^2}{2eB}\biggr) \\
  &\qquad -2\ln\biggl(\frac{\mu_N+\sqrt{\mu_N^2 - m^2}}{m}\biggr)
   + 2\sqrt{1-\frac{m^2}{\mu_N^2}}\; \Bigg]\;,
  \end{split}
\label{eq:rot-proper}
\end{equation}
where $\EM$ is the Euler-Mascheroni constant, $\Gamma(z)$ denotes the
gamma function, and $\Lpt$ stands for the cutoff parameter in the
proper-time regularization.  In this gap
equation~\eqref{eq:rot-proper}, the terms in the third line result from the
$n=0$ mode in Eq.~\eqref{eq:FOmega2}.
We can find $\Omega_c$ from the above gap equation with $m\to 0$
substituted, and the analytical result is
\begin{equation}
  \begin{split}
    \Omega_c(eB) &=  \frac{\sqrt{\pi}}{S\sqrt{eB}} \exp\biggl[-\frac{2\pi^2}
    {eB}\biggl(\frac{1}{G}-\frac{1}{G_c}\biggr) + 1 \biggr]\\
    &\simeq \frac{1.53\times 10^{-6}}{\sqrt{eB}} \exp\biggl(-\frac{0.610\Lambda^2}{eB}
     \biggr) \;,
  \end{split}
\label{eq:Omegac-analytical}
\end{equation}
where $G_c = 4\pi/\Lpt^2$ is the critical coupling for $\Omega=\sqrt{eB}=0$
that is found in the proper-time regularization.  In the second line
in Eq.~\eqref{eq:Omegac-analytical}, we utilized the parameters of
Eqs.~\eqref{eq:Gc}, \eqref{eq:weakG} and \eqref{eq:parameters}.  On the other hand, we
can numerically evaluate $\Omega_c$ as a function of $eB$ as displayed
in Fig.~\ref{fig:Omegac}.  From the linearity in Fig.~\ref{fig:Omegac}
the numerical fit leads to
\begin{equation}
 \Omega_c(eB) \simeq \frac{1.58\times 10^{-6}}{\sqrt{eB}} \exp\biggl(
  -\frac{0.609\Lambda^2}{eB} \biggr) \;. 
\label{eq:Omegac-numerical}
\end{equation}
This fitting result ensures that Eq.~\eqref{eq:integral} is a good
approximation for the parameters in Eq.~\eqref{eq:parameters}.


\subsection{Dynamical mass at strong coupling ($G>G_c$)}

\begin{figure}
  \begin{center}
    \includegraphics[width=1\columnwidth]{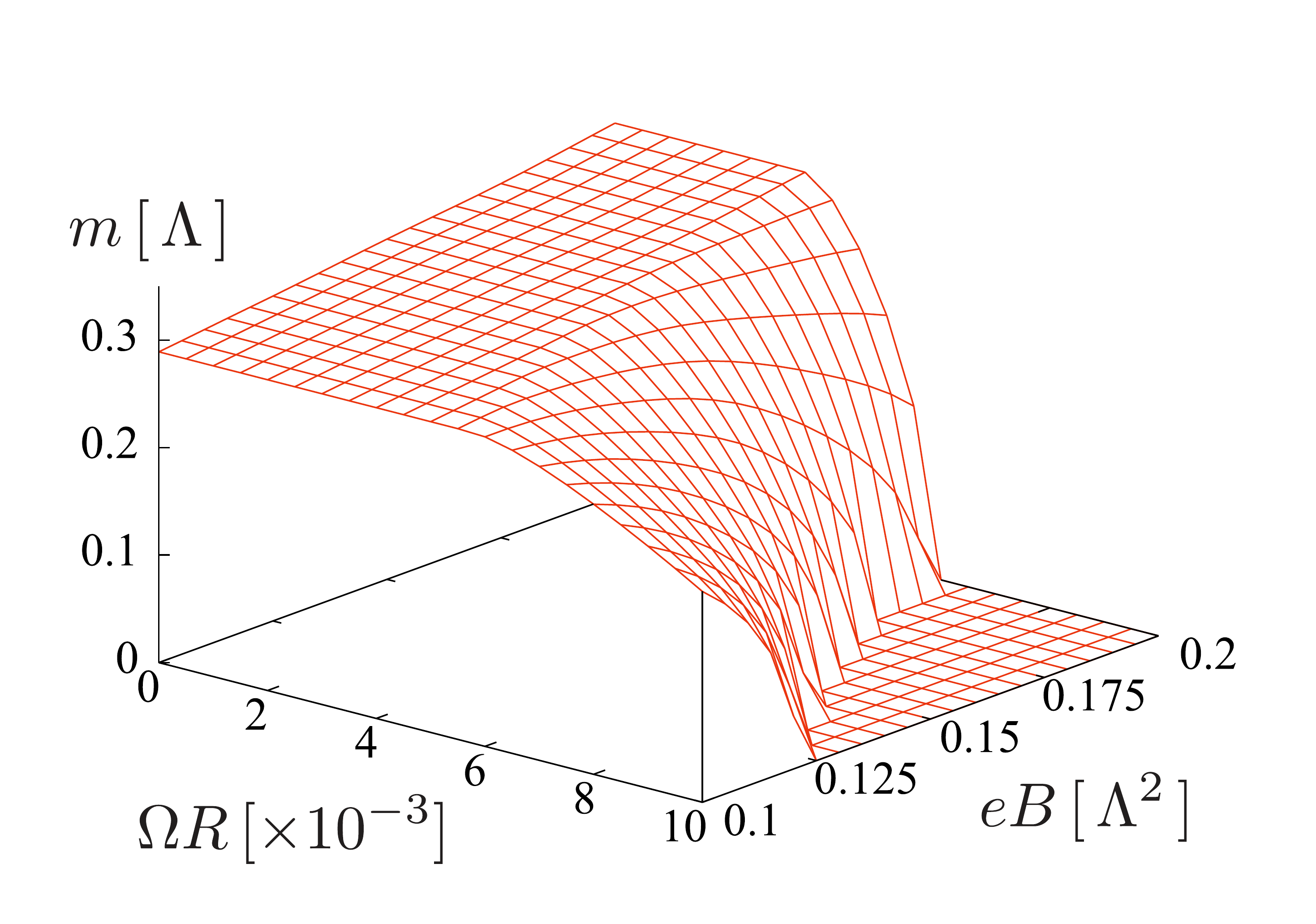}
    \caption{3D plot for the dynamical mass as a function of $\Omega$
    and $eB$ at strong coupling.  For large $\Omega$, chiral symmetry
    is restored by $eB$, which manifests the inverse magnetic catalysis
    or the rotational magnetic inhibition.}
    \label{fig:3dstrong}
  \end{center}
\end{figure}

We shall next focus on the following strong region:
\begin{equation}
  G=1.11 G_c\;.
\label{eq:strongG}
\end{equation}
We note that dynamically determined $m$ with
the above strong-coupling is about 20 times larger than
$m_{\text{dyn}}$ at weak coupling.  We show the numerical results in
Fig.~\ref{fig:3dstrong}.  Below are several remarks on the results.

\vspace{1em}

(I) For small
angular velocity, the dynamical mass is almost independent of $\Omega$
and $eB$.  With increasing $\Omega$ the dynamical mass is eventually
suppressed by larger magnetic field, i.e.\ a counterpart of the
finite-density inverse magnetic catalysis is manifested.  We would
call this decreasing behavior of the mass for larger magnetic field
the ``rotational magnetic inhibition'' in this paper.  In
Fig.~\ref{fig:3dstrong}  we see that the dynamical mass starts to
drop around $\mu_N = \Omega N \sim \sqrt{eB}$.  The same is true for the
finite-density inverse magnetic catalysis observed around
$\mu \sim \sqrt{eB}$~\cite{Preis:2010cq}.

We notice that there is a drastic difference between the weak and the
strong coupling results and this difference is attributed to higher
Landau levels relevant for the determination of the dynamical mass.
In the weak coupling case only a small number of the Landau levels
contribute to the gap equation, while many more Landau levels get
involved as the coupling constant becomes larger.  This is essential
for the realization of the rotational magnetic inhibition as well as of
the inverse magnetic catalysis at finite density.
\vspace{0.5em}

(II) Let us now take a close look at some possible difference between the
finite-density inverse magnetic catalysis and the rotational magnetic
inhibition.  The QCD vacuum has a rich content with $\mu$ and $eB$,
and for $G>G_c$, particularly, the de~Haas-van~Alphen
oscillation~\cite{de1930dependence,*de1930oscillations} may lead to
several local minima of the gap equation~\cite{Ebert:1999ht}.
However, Fig.~\ref{fig:3dstrong} shows that this is not the case for
the rotational magnetic inhibition.  To see the microscopic details, we
should clarify the profile of the following function:
\begin{equation}
  F(m) = \frac{1}{G} - \frac{1}{\pi}(F_0-F_\Omega)\;.
\label{eq:F}
\end{equation}
This function itself is continuous for any $m$, but the derivative is
not, that is,
\begin{equation}
  \begin{split}
    \frac{dF(m)}{dm} =  &\frac{m}{\pi S} \sum_{n=0}^\infty \alpha_n
    \sum_{\ell=-n}^{N-n}\Biggl[ \int_0^\infty \frac{dp_z\,f(p_z,n;\Lambda)}
    {(p_z^2 + m_n^2)^{3/2}} \\
    & - \int_0^{k_{nj}} \frac{dp_z\,f(p_z,n;\Lambda)}{(p_z^2 + m_n^2)^{3/2}}
    \theta (\Omega |j| - m_n)  \\
    & - \frac{\,f(\Omega|j|,0;\Lambda)}{\Omega |j|\sqrt{(\Omega j)^2-m_n^2}}
    \theta(\Omega |j| - m_n)
    \Biggr]\;,
  \end{split}
\label{eq:dF}
\end{equation}
which negatively diverges at
\begin{equation}
 m = \sigma_{nj} \equiv \sqrt{(\Omega j )^2 - 2neB}\;.
\end{equation}
If $m$ is greater than $\sigma_{nj}$, only the first term in the
right-hand side of Eq.~\eqref{eq:dF} remains nonvanishing for a fixed
$n$, and we can confirm that Eq.~\eqref{eq:dF} turns out to be
positive.  Thus, we find,
\begin{equation}
  \frac{dF(m)}{dm}\bigg|_{m\to \sigma_{nj}-0} \!\!\!= -\infty\;,
  \qquad
  \frac{dF(m)}{dm}\bigg|_{m\to \sigma_{nj}+0} \!\!\!> 0\;,
\end{equation}
and so $F(m)$ may not be a monotonic function.  This behavior in the
vicinity of $m = \sigma_{nj}$ might cause technical difficulties to
deal with multiple zeros of $F(m)$.  Indeed, a simple replacement of
$\Omega j$ with $\mu$ leads to a profile of $F(m)$ having the
de~Haas-van~Alphen oscillation for the mass
gap~\cite{Ebert:1999ht,Preis:2012fh}.  In the case with rotation,
however, these singularities are not very important for the solution
of the gap equation.  In our numerical studies, actually, we find that
$F(m)$ has only one solution, which is explained as follows.

If the effective chemical potential is not too large, i.e.\
$\mu_N =\Omega N \lesssim \Lambda$, we can practically remove the
cutoff function from the third term in Eq.~\eqref{eq:dF}.  Then we can
approximate the $\ell$-sum in the third term by an integration as we
did in Eq.~\eqref{eq:integral}:
\begin{equation}
  \begin{split}
    &\sum_{\ell = -n}^{N-n}\frac{1}{\Omega |j|\sqrt{(\Omega j)^2-m_n^2}}
    \,\theta(\Omega |j|- m_n) \\
    & \simeq \frac{1}{\Omega} \int_{m_n}^{\mu_N}\frac{d\mu}
    {\mu\sqrt{\mu^2-m_n^2}}\,\theta(\mu_N- m_n)  \\
    & =  \frac{1}{\Omega m_n}\biggl[ \frac{\pi}{2} - \tan^{-1}
    \biggl(\frac{m_n}{\sqrt{\mu_N^2-m_n^2}}\biggr) \biggr]
    \,\theta(\mu_N- m_n)\;,
  \end{split}
\label{eq:divergence}
\end{equation}
which is finite even at $m = \sigma_{nN}$.  In the opposite limit of
large $\mu_N >  \Lambda$, we can make a similar argument with $\mu_N$
replaced with $\Lambda$ to confirm that no singularity appears from an
approximated form of Eq.~\eqref{eq:dF}.  Therefore, it is effectively
possible for us to regard $F(m)$ as a monotonically increasing
function in our numerical analysis.  This explains why
Fig.~\ref{fig:3dstrong} does not show the de~Haas-van~Alphen
oscillation.
\vspace{0.5em}

(III) We briefly discuss the physical implications of the rotational
magnetic inhibition to realistic situations at strong coupling (with
chiral symmetry breaking in the vacuum).  First, let us take an
example from the condensed matter physics system;  for a material with
$R = 1\,\text{cm}$ (e.g.\ graphene or 3D Dirac semimetal) under the
magnetic field $B = 1.7\times 10^6\,\text{G}$, we find that the
rotational magnetic inhibition takes place around
$\mu_N\sim \sqrt{eB}$, that is,
$\Omega \sim \sqrt{eB}v_F/N \simeq 2.5 \times 10^2\, \text{s}^{-1}$
where we adopt $v_F = 10^6\,\text{m/s}$ from the Fermi velocity of the
quasiparticles in graphene~\cite{Novoselov:2005kj}.  This suggests
that the rotational magnetic inhibition should be an observable effect
in a table-top experiment.

Another interesting environment where the rotational magnetic
inhibition could be activated is the neutron star. If one makes a
na\"{i}ve estimate for a millisecond pulsar,
$\Omega\sim 10^3\;\text{s}^{-1}$ and $R\sim 10^4\;\text{m}$ would lead
to $\Omega R\sim \mathcal{O}(10^{-1})$.  In view of
Fig.~\ref{fig:3dstrong} one may well conclude that chiral symmetry
should be restored at small $eB$ or even at zero $eB$.  This is a very
fascinating possibility that might have an impact to the construction
of the equation of state (EoS);  the neutron star EoS could get harder
thanks to the rotational magnetic inhibition.  We have to leave
quantitative studies for future works, however, because
$R\simeq 10^4\;\text{m}$ is outside of the region of $R$ that we
adopted in this paper.  Here we just emphasize that due to the 
quadratic dependence, $\mu_N \propto R^2$, the chiral condensate with 
larger $R$ is generally more sensitive to the rotational effect. Thus, 
the rotational magnetic inhibition must be definitely a sizable effect 
for the neutron star physics.



\section{Conclusion and outlook}

We analyzed the Dirac equation with magnetic field background in a
rotating frame.  We showed that rotation should modify the Landau
levels and resolve the Landau degeneracy.  As a result, rotation plays
a role similar with finite chemical potential.  In the weak coupling
case where chiral symmetry is not yet broken in the vacuum, the
dynamical mass is induced by the magnetic field according to the
well-known magnetic catalysis.  Together with rotation, we found that
the dynamical mass is suppressed with increasing angular velocity
$\Omega$.  For strong coupling case that realizes chiral symmetry
breaking in the vacuum, we discovered opposite behavior of the
dynamical mass.  At finite $\Omega$ (like finite density), the
dynamical mass decreases with increasing magnetic field.  This
phenomenon is an inverse of the conventional magnetic catalysis, and
could be regarded as an example of the inverse magnetic catalysis.  To
distinguish our finding from the inverse magnetic catalysis at finite
density or finite temperature, we named this novel phenomenon the
``rotational magnetic inhibition.''

We should note that the rotational responses obtained from our 
calculation are model-independent. We introduced a specific cutoff 
function for the $p_z$-integral and the $n$-sum to evaluate the chiral 
condensate within the NJL model. Nevertheless, we also confirmed that the 
behaviors of the chiral condensate are qualitatively unchanged even if 
we use other parameters. Additionally once we regularize the $n$-sum, 
the upper bound of angular momentum $\ell$ is automatically determined 
by the Landau degeneracy factor, which is irrelevant to model 
parameters. Thus whatever model and parameters we adopt, rotation should 
give density-like contributions to the dynamical mass, as expected in the 
level of the energy spectrum.

We can find many possible applications of the rotational magnetic
inhibition.  We could discuss it in condensed matter systems, in the
cores of the neutron star, and in noncentral relativistic heavy-ion
collision experiments~\cite{Csernai:2011gg}.  We gave a rough estimate
of whether our analysis could be relevant for those systems.  For the
heavy-ion collision the estimate of $\Omega$ is still unclear and it
is difficult to make any decisive statement.  For the neutron star, 
although angular velocity itself is much smaller than the QCD scale, 
it is obvious from our results that the rotation should give a sizable 
modification to the dynamical mass through an effective chemical 
potential $\mu_N=\Omega N$. Hence we must consider the rotational 
influence to the equation of state (EoS);  so far, its angular 
velocity seems too small to affect the QCD dynamics and the rotation has
been treated only as a global effect in the EoS~\cite{Cook:1993qr} (see also Refs.~\cite{friedman1986rapidly,%
glendenning1997signal,chubarian1999deconfinement,haensel2007neutron}). 
We are now making a progress to go beyond such a treatment.

To this end, though it is beyond our scope of the present paper, we
would need to consider spatially inhomogeneous condensates as done in
finite-density systems (see Ref.~\cite{Buballa:2014tba} and references
therein) together with a magnetic
field~\cite{Fukushima:2013zga,Carignano:2015kda} and also in systems
influenced by surrounding geometrical
effects~\cite{Flachi:2011sx,*Flachi:2013iia,*Flachi:2014jra,Benic:2015qha}.
Most probably the chiral condensate could be decreasing as the radial
distance from the rotation center becomes larger.  In the present
analysis we assumed a homogeneous condensate using not the effective
action but the effective potential only.  This is how the results
depend on the bulk parameter $R$, which might be augmented with local
radial distance dependence.  The main purpose of this current study is
to pursue the analogy between rotation and density and we would leave
more quantitative discussions including spatial inhomogeneity for
future works.


\acknowledgments
This work was supported by Japan Society for the Promotion of 
Science (JSPS) KAKENHI Grant No.\ 15H03652 and
15K13479 (K.~F.), JSPS Research Fellowships for Young Scientists
(K.~M.), and Shanghai Natural Science Foundation (Grant No. 14ZR1403000) and the Young 1000 Talents Program of China (H.-L.~C. and X.-G.~H).

\appendix
\section{Dirac equation in a rotating frame}
\label{app:solution}
We solve the Dirac equation in a rotating frame \eqref{eq:Deq}. We find 
that the quantum number $\ell$ in Eq.~\eqref{eq:DR} is same as
that without rotation. We also show that a quite large system size as 
Eq.~\eqref{eq:condition} is necessary for the Landau quantization.

We rewrite Eq.~\eqref{eq:Deq} as the following equation:
\begin{equation}
 \begin{split}
  0 &= [i\gamma^\mu (D_\mu + \Gamma_\mu ) + m ][i\gamma^\mu (D_\mu +
  \Gamma_\mu ) - m ]\psi \\
  & = \bigg[(i\partial_t + \Omega \hat L_z + \Omega \sigma^{12})^2
  +\partial_x^2 + \partial_y^2 + \partial_z^2  \\
  & + eB(\hat L_z + 2\sigma^{12}) - \left( \frac{eB}{2}\right)^2
  (x^2+y^2)  -m^2 \bigg]\psi\\
  & = \bigg[(i\partial_t - i \Omega \partial_\theta + \Omega
  \sigma^{12})^2 + \partial_r^2 + \frac{1}{r}\partial_r + \frac{1}{r^2}
  \partial_\theta^2  \\
  &+ eB(-i\partial_\theta + 2\sigma^{12}) - \left( \frac{eB}{2}
  \right)^2r^2 + \partial_z^2   -m^2 \bigg]\psi\;.
 \end{split}
\end{equation}
The last line is written with the cylindrical coordinate $x^\mu =
(t,r,\theta,z)$. Taking the chiral representation
\begin{equation}
 \psi =
 \begin{pmatrix}
  \chi \\
  \varphi
 \end{pmatrix}
\;,
\label{eq:Drep}
\end{equation}
the solution is written as the following function:
\begin{equation}
\chi = e^{-iEt+ip_z z}
\begin{pmatrix}
e^{i\ell_+\theta}\chi_+(r) \\
e^{i\ell_-\theta}\chi_-(r)
\end{pmatrix}
\;,
\end{equation}
where $\ell_\pm$ is an integer and the radial function is defined as 
$\sigma^3\chi_\pm = \pm\chi_\pm$. In what follows, we solve only the 
equation of $\chi$, but that of $\varphi$ can be solved in the same way. 
Because of the rotational invariance, $\psi$ should also be an 
eigenfunction for $\hat J_z = \hat L_z + \sigma^{12}$. In other words, 
$\chi_\pm$ have  equivalent total angular momenta, which leads
\begin{equation}\label{eq:lpm}
\ell_+ = \ell_- -1\;.
\end{equation}
We note that at this stage, there are no constraint for $\ell_\pm$
other than Eq.~\eqref{eq:lpm}.

Substituting $\psi$ with Eq.~\eqref{eq:Drep}, we obtain the equation for
$\chi_\pm$:
\begin{equation}
 \begin{split}
  &\biggl[\Bigl\{E + \Omega (\ell_\pm \pm 1/2)\Bigr\}^2 - p_z^2 -m^2 +
  eB(\ell_\pm \pm 1) \\
  & \quad \quad + \partial_r^2 + \frac{1}{r}\partial_r
  -\frac{\ell_\pm^2}{r^2} - \left( \frac{eB}{2}\right)^2 r^2
   \biggr]\chi_\pm = 0\;.
 \end{split}
\end{equation}
The solution for this equation is written with the confluent 
hypergeometric function:
\begin{equation}
  \begin{split}
    \chi_\pm &= r^{|\ell_\pm|}e^{-eBr^2/4}\Bigl[c_1 M(a,|\ell_\pm|+1,
    eBr^2/2) \\
    &+ c_2 (eBr^2/2)^{-|\ell_\pm|} M(a-|\ell_\pm|, 1-|\ell_\pm|,
    eBr^2/2)\Bigr]
  \end{split}
\end{equation}
with 
\begin{equation}\label{eq:a}
 \begin{split}
  a &= \frac{1}{2}(|\ell_\pm| \mp 1 - \ell_\pm + 1)\\
  &\quad -\frac{1}{2eB}\bigg[ \Big\{E+\Omega(\ell_\pm \pm 1/2)\Big\}^2
  -p_z^2 -m^2\bigg]\;.
 \end{split}
\end{equation}
It is necessary that $\chi_\pm$ be finite at arbitrary $r$ because 
of normalizability. The finiteness of
$\chi_\pm(r\to 0)$ demands $c_2=0$. Also to keep
$\chi_\pm(r\to\infty)$ finite, $a$ should be a nonpositive integer:
\begin{equation}\label{eq:np}
-a \equiv n_p = 0,1,2,\cdots\;.
\end{equation}
In this case, this hypergeometric function is reduced to an associated
Laguerre polynomial:
\begin{equation}
 M(-n_p, |\ell_\pm| +1, eBr^2/2) = C_{n,|\ell_\pm|} L_{n_p}
 ^{|\ell_\pm|}(eBr^2/2)\;,
\end{equation}
where $C_{n,|\ell_\pm|}$ is a constant independent of $r$.  From
Eq.~\eqref{eq:np}, we find that the energy dispersion relation is
quantized:
\begin{equation}\label{eq:DRapp}
\begin{split}
&\Big[E+\Omega(\ell_\pm \pm 1/2)\Big]^2 \\
&= eB(2n_p +|\ell_\pm| -\ell_\pm +1\mp 1) +p_z^2 +m^2\;.
\end{split}
\end{equation}
This dispersion can be reduced to Eq.~\eqref{eq:DR}
if we introduce the new integers defined by
\begin{equation}
\begin{split}
n &= n_+ = n_-+1, \\
n_\pm &\equiv n_p + \frac{1}{2}(|\ell_\pm| -\ell_\pm + 1 \mp 1 )\;.
\end{split}
\label{eq:n}
\end{equation}
We note that these equations imply the lower bounds for $\ell_\pm$:
\begin{equation}
\ell\equiv \ell_+ = \ell_- - 1 \geq -n\;.
\end{equation} 
Finally using the property of the Laguerre function,
\begin{equation}
 L_{n+\ell}^{-\ell}(x^2) \propto x^{2\ell} L_{n}^{\ell}(x^2)
 \quad (\,\text{for\ } \ell\leq 0 \,)\;,
\end{equation}
we obtain the eigenfunction as the following simpler form:
\begin{equation}
 \begin{split}
  \chi_{n,\ell,\pm}(r) \propto r^{\ell}e^{-eBr^2/4} L_{n}^{\ell}
  (eBr^2/2)\;.
 \end{split}
\label{eq:eigenfunction}
\end{equation}

We mention that the above argument is valid even in the 
$\Omega = 0$ case. This means that the quantum number $\ell$ is 
the same as the one without rotation. Therefore the possible range of
$\ell$ in a rotating frame is same as the one without rotation, i.e., Eq.~\eqref{eq:lrange} (see Appendix~\ref{app:degeneracy}).

For the discussion in this paper, it is significant that the
quantization in terms of $n_p$ in Eq.~\eqref{eq:np} is
performed only if the wave function converges at infinity.
The same is true for the quantization of general harmonic
oscillators. This is why we need to
consider systems with a much larger radius than $\ell_B =
1/\sqrt{eB}$, as shown in Eq.~\eqref{eq:condition}.

\section{Landau degeneracy in cylindrers}\label{app:degeneracy}
Based on the Klein$-$Gordon equation for charged scalar in external
magnetic field, we briefly show that even in cylindrical coordinate
the Landau degeneracy factor is given by
Eq.~\eqref{eq:degeneracy}, as well as in Cartesian coordinate. It is
also proved that the range of quantum number $\ell$ for the $n$th
Landau level is given by Eq.~\eqref{eq:lrange}.

\subsection{Landau quantization for general gauges}
We prepare the Landau quantization for general gauge.
The Klein-Gordon equation in a magnetic field ${\boldsymbol B} =
B\hat z$ is as follows: 
\begin{equation}
\begin{split}
(\partial_t^2 - \partial_z^2 + m^2 - D_x^2 - D_y^2 ) \Phi = 0\;.
\end{split}
\end{equation}
It is clear that the solution is the form given by $\Phi =
e^{-i\varepsilon t+ ik z}\phi(x,y)$. The Klein$-$Gordon equation is
then reduced to
\begin{equation}\label{eq:eigeneq}
 \hat H\phi(x,y) = \lambda \phi(x,y)
\end{equation}
with
\begin{equation}
 \hat H \equiv -(D_x^2 + D_y^2)\;,\quad \lambda \equiv \varepsilon^2
 - k^2 - m^2\;.
\end{equation}
This eigenvalue equation can be solved by introducing the ladder
operators:
\begin{equation}
 a \equiv \frac{i}{\sqrt{2eB}}(D_x + iD_y)\;,\quad a^\dag \equiv
 \frac{i}{\sqrt{2eB}}(D_x - iD_y)\;,
\end{equation}
which satisfy $[a,a^\dag] = 1$. Therefore the eigenstates for
Eq.~\eqref{eq:eigeneq} are $|n\rangle \propto (a^\dag)^n|0\rangle$,
and the corresponding eigenvalue is given by
\begin{equation}
\begin{split}
\hat H | n\rangle = eB(2n + 1)|n\rangle\;.
\end{split}
\end{equation}

\subsection{Landau quantization for symmetric gauge}

When we utilize the cylindrical coordinate, the symmetric gauge 
$A_\mu =(0,By/2,-Bx/2,0)$ is most useful because the Hamiltonian in
Eq.~\eqref{eq:eigeneq} respects the rotational symmetry. Instead of
$(r,\theta)$, we use the complex coordinate defined by
\begin{equation}
z \equiv x + iy\;,\quad \bar z \equiv x-iy\;.
\end{equation}
We also introduce the new notations for the derivatives in terms of
$z$ and $\bar z$; $\partial \equiv \partial/\partial z$ and
$\bar\partial \equiv \partial/\partial\bar z$. Then the ladder
operators are also rewritten by $(z,\hat z)$:
\begin{equation}
 \begin{split}
  a = \frac{-i}{\sqrt{2eB}}\left(2\bar\partial +\frac{eB}{2} z\right)\;,
  \quad a^\dag = \frac{-i}{\sqrt{2eB}}\left(2\partial -\frac{eB}{2}
  \bar z\right)\;.
 \end{split}
\end{equation}
The ground state is obtained by the condition $a|0\rangle = 0$:
\begin{equation}
  \langle z,\bar z | a | 0 \rangle = -i \frac{1}{\sqrt{2eB}} \left(
  2\bar\partial +\frac{eB}{2} z\right) \phi(z,\bar z) = 0\;.
\end{equation}
The solution is given by
\begin{equation}
\begin{split}
\phi(z,\bar z) = \tilde \phi(z) e^{-eBz\bar z / 4}\;,
\end{split}
\end{equation}
where $\tilde \phi(z)$ denotes a function of $z$. In principle, we have
no condition for the choice of $\tilde \phi(z)$, except for the
analyticity. This facultativity of $\tilde \phi(z)$ comes from assuming 
an infinitely large system in our calculation. In other words, the 
eigenvalue equation of the harmonic oscillator cannot be solved in 
finite-size systems (see Appendix~\ref{app:solution}).

In order to find $\tilde \phi(z)$,
we analyze another quantum number, i.e., the canonical angular momentum. 
Because of the rotational-invariance of the Hamiltonian, the eigenstate 
of the Hamiltonian can be also the eigenstate of the angular momentum 
$\hat L_z = xp_y - yp_x = z\partial- \bar z\bar \partial$. Let us 
introduce the new ladder operators:
\begin{equation}
 \begin{split}
  b \equiv \frac{1}{\sqrt{2eB}} \left(2\partial + \frac{eB}{2} \bar
  z\right),\quad b^\dag \equiv \frac{1}{\sqrt{2eB}} \left(-2\bar
  \partial+\frac{eB}{2} z\right)\;,
 \end{split}
\end{equation}
which satisfy $[b,b^\dag] = 1$. We represent the angular momentum as
the ladder operators:
\begin{equation}
\hat L_z  =b^\dag b - a^\dag a\;.
\end{equation}
We define the simultaneous eigenstates for $a^\dag a$ and $b^\dag b$:
\begin{equation}
 \begin{split}
  a^\dag a | n, n_p \rangle = n | n, n_p \rangle,\quad b^\dag b
  | n, n_p \rangle = n_p | n, n_p \rangle\;,
 \end{split}
\end{equation}
for $n, n_p = 0,1,\cdots$. Instead of $n_p$, we designate these 
eigenstates by the new quantum number $\ell = n_p-n$:
\begin{equation}
 \begin{split}
  & \langle z,\bar z | n, n_p \rangle = \phi_{nn_p}(z,\bar z )
  \equiv \psi_{n\ell} (z,\bar z )\;,\\
  &\hat L_z\psi_{n\ell} = (b^\dag b - a^\dag a) \psi_{n\ell} =
  \ell \psi_{n\ell}\;. \\ \\
 \end{split}
\end{equation}
We note that the non-negativities of $n$ and $n_p$ lead to the lower
bound of $\ell$:
\begin{equation}
\ell\geq -n\;.
\end{equation}

We produce a ground state by the operation of ladder operator
$b^\dag$:
\begin{equation}\label{eq:gs}
 \begin{split}
  \psi_{0\ell}(z,\bar z) &\propto (b^\dag)^\ell
  \psi_{00}(z,\bar z) \\
  &\propto z^\ell e^{-eBz\bar z /4}\;. 
 \end{split}
\end{equation}
From this eigenstate, we find that $\ell$ corresponds to the 
degenerate quantum number, which is irrelevant to the energy level. 
Thus the possible range of $\ell$ is nothing but the Landau 
degeneracy factor. In order to calculate the degeneracy factor, we 
focus on the following equation:
\begin{equation}\label{eq:req}
\begin{split}
 \frac{d}{dr} (2\pi r|\psi_{0\ell}|^2) = 0\;, \\  
\end{split}
\end{equation}
which determines the radius that gives the maximum value of this 
distribution.
If we consider the system to be the cylinder with radius $R$, the
solution for Eq.~\eqref{eq:req} should be smaller than $R$, which
leads to the upper bound of $\ell$:
\begin{equation}
\begin{split}
\ell \leq \frac{eBR^2}{2} = \frac{eBS}{2\pi}\;.
\end{split}
\end{equation}
Therefore from this upper bound and the lower bound $\ell\geq -n = 0$, the
degeneracy factor in the cylindrical coordinate is given by
Eq.~\eqref{eq:degeneracy}.

Higher excited states with $n\geq 1$ are calculated in a similar way
to ground states:
\begin{equation}\label{eq:es}
 \begin{split}
  \psi_{n\ell} (z,\bar z)
  &\propto(a^\dag)^n(b^\dag)^{n+\ell} \psi_{00}
  (z,\bar z) \\
  & \propto z^\ell e^{-eBz\bar z /4} L^\ell_n (eBz\bar z /2)\;,
 \end{split}
\end{equation}
which is same as Eq.~\eqref{eq:eigenfunction} if we use $z=re^{i\theta}$
and $\bar z=re^{-i\theta}$.
The upper bound of $\ell$ for exited states cannot directly be found
from Eq.~\eqref{eq:es} while the one for the ground state is derived
from the wave function \eqref{eq:gs}. Nevertheless the upper bound of
$\ell$ for exited states is obviously $N-n$ because the degeneracy factor
$N$ is a common quantity for all Landau levels. From this and the lower
bound $\ell \geq -n$, we obtain Eq.~\eqref{eq:lrange}.

\bibliography{ref}

\begin{thebibliography}{88}%
\makeatletter
\providecommand \@ifxundefined [1]{%
 \@ifx{#1\undefined}
}%
\providecommand \@ifnum [1]{%
 \ifnum #1\expandafter \@firstoftwo
 \else \expandafter \@secondoftwo
 \fi
}%
\providecommand \@ifx [1]{%
 \ifx #1\expandafter \@firstoftwo
 \else \expandafter \@secondoftwo
 \fi
}%
\providecommand \natexlab [1]{#1}%
\providecommand \enquote  [1]{``#1''}%
\providecommand \bibnamefont  [1]{#1}%
\providecommand \bibfnamefont [1]{#1}%
\providecommand \citenamefont [1]{#1}%
\providecommand \href@noop [0]{\@secondoftwo}%
\providecommand \href [0]{\begingroup \@sanitize@url \@href}%
\providecommand \@href[1]{\@@startlink{#1}\@@href}%
\providecommand \@@href[1]{\endgroup#1\@@endlink}%
\providecommand \@sanitize@url [0]{\catcode `\\12\catcode `\$12\catcode
  `\&12\catcode `\#12\catcode `\^12\catcode `\_12\catcode `\%12\relax}%
\providecommand \@@startlink[1]{}%
\providecommand \@@endlink[0]{}%
\providecommand \url  [0]{\begingroup\@sanitize@url \@url }%
\providecommand \@url [1]{\endgroup\@href {#1}{\urlprefix }}%
\providecommand \urlprefix  [0]{URL }%
\providecommand \Eprint [0]{\href }%
\providecommand \doibase [0]{http://dx.doi.org/}%
\providecommand \selectlanguage [0]{\@gobble}%
\providecommand \bibinfo  [0]{\@secondoftwo}%
\providecommand \bibfield  [0]{\@secondoftwo}%
\providecommand \translation [1]{[#1]}%
\providecommand \BibitemOpen [0]{}%
\providecommand \bibitemStop [0]{}%
\providecommand \bibitemNoStop [0]{.\EOS\space}%
\providecommand \EOS [0]{\spacefactor3000\relax}%
\providecommand \BibitemShut  [1]{\csname bibitem#1\endcsname}%
\let\auto@bib@innerbib\@empty
\bibitem [{\citenamefont {Cheng}\ and\ \citenamefont
  {Olinto}(1994)}]{Cheng:1994yr}%
  \BibitemOpen
  \bibfield  {author} {\bibinfo {author} {\bibfnamefont {B.-l.}\ \bibnamefont
  {Cheng}}\ and\ \bibinfo {author} {\bibfnamefont {A.~V.}\ \bibnamefont
  {Olinto}},\ }\href {\doibase 10.1103/PhysRevD.50.2421} {\bibfield  {journal}
  {\bibinfo  {journal} {Phys. Rev.}\ }\textbf {\bibinfo {volume} {D50}},\
  \bibinfo {pages} {2421} (\bibinfo {year} {1994})}\BibitemShut {NoStop}%
\bibitem [{\citenamefont {Baym}\ \emph {et~al.}(1996)\citenamefont {Baym},
  \citenamefont {Bodeker},\ and\ \citenamefont {McLerran}}]{Baym:1995fk}%
  \BibitemOpen
  \bibfield  {author} {\bibinfo {author} {\bibfnamefont {G.}~\bibnamefont
  {Baym}}, \bibinfo {author} {\bibfnamefont {D.}~\bibnamefont {Bodeker}}, \
  and\ \bibinfo {author} {\bibfnamefont {L.~D.}\ \bibnamefont {McLerran}},\
  }\href {\doibase 10.1103/PhysRevD.53.662} {\bibfield  {journal} {\bibinfo
  {journal} {Phys. Rev.}\ }\textbf {\bibinfo {volume} {D53}},\ \bibinfo {pages}
  {662} (\bibinfo {year} {1996})},\ \Eprint
  {http://arxiv.org/abs/hep-ph/9507429} {arXiv:hep-ph/9507429 [hep-ph]}
  \BibitemShut {NoStop}%
\bibitem [{\citenamefont {Grasso}\ and\ \citenamefont
  {Rubinstein}(2001)}]{Grasso:2000wj}%
  \BibitemOpen
  \bibfield  {author} {\bibinfo {author} {\bibfnamefont {D.}~\bibnamefont
  {Grasso}}\ and\ \bibinfo {author} {\bibfnamefont {H.~R.}\ \bibnamefont
  {Rubinstein}},\ }\href {\doibase 10.1016/S0370-1573(00)00110-1} {\bibfield
  {journal} {\bibinfo  {journal} {Phys. Rept.}\ }\textbf {\bibinfo {volume}
  {348}},\ \bibinfo {pages} {163} (\bibinfo {year} {2001})},\ \Eprint
  {http://arxiv.org/abs/astro-ph/0009061} {arXiv:astro-ph/0009061 [astro-ph]}
  \BibitemShut {NoStop}%
\bibitem [{\citenamefont {Duncan}\ and\ \citenamefont
  {Thompson}(1992)}]{Duncan:1992hi}%
  \BibitemOpen
  \bibfield  {author} {\bibinfo {author} {\bibfnamefont {R.~C.}\ \bibnamefont
  {Duncan}}\ and\ \bibinfo {author} {\bibfnamefont {C.}~\bibnamefont
  {Thompson}},\ }\href {\doibase 10.1086/186413} {\bibfield  {journal}
  {\bibinfo  {journal} {Astrophys. J.}\ }\textbf {\bibinfo {volume} {392}},\
  \bibinfo {pages} {L9} (\bibinfo {year} {1992})}\BibitemShut {NoStop}%
\bibitem [{\citenamefont {Skokov}\ \emph {et~al.}(2009)\citenamefont {Skokov},
  \citenamefont {Illarionov},\ and\ \citenamefont {Toneev}}]{Skokov:2009qp}%
  \BibitemOpen
  \bibfield  {author} {\bibinfo {author} {\bibfnamefont {V.}~\bibnamefont
  {Skokov}}, \bibinfo {author} {\bibfnamefont {A.~{\relax Yu}.}\ \bibnamefont
  {Illarionov}}, \ and\ \bibinfo {author} {\bibfnamefont {V.}~\bibnamefont
  {Toneev}},\ }\href {\doibase 10.1142/S0217751X09047570} {\bibfield  {journal}
  {\bibinfo  {journal} {Int. J. Mod. Phys.}\ }\textbf {\bibinfo {volume}
  {A24}},\ \bibinfo {pages} {5925} (\bibinfo {year} {2009})},\ \Eprint
  {http://arxiv.org/abs/0907.1396} {arXiv:0907.1396 [nucl-th]} \BibitemShut
  {NoStop}%
\bibitem [{\citenamefont {Voronyuk}\ \emph {et~al.}(2011)\citenamefont
  {Voronyuk}, \citenamefont {Toneev}, \citenamefont {Cassing}, \citenamefont
  {Bratkovskaya}, \citenamefont {Konchakovski},\ and\ \citenamefont
  {Voloshin}}]{Voronyuk:2011jd}%
  \BibitemOpen
  \bibfield  {author} {\bibinfo {author} {\bibfnamefont {V.}~\bibnamefont
  {Voronyuk}}, \bibinfo {author} {\bibfnamefont {V.~D.}\ \bibnamefont
  {Toneev}}, \bibinfo {author} {\bibfnamefont {W.}~\bibnamefont {Cassing}},
  \bibinfo {author} {\bibfnamefont {E.~L.}\ \bibnamefont {Bratkovskaya}},
  \bibinfo {author} {\bibfnamefont {V.~P.}\ \bibnamefont {Konchakovski}}, \
  and\ \bibinfo {author} {\bibfnamefont {S.~A.}\ \bibnamefont {Voloshin}},\
  }\href {\doibase 10.1103/PhysRevC.83.054911} {\bibfield  {journal} {\bibinfo
  {journal} {Phys. Rev.}\ }\textbf {\bibinfo {volume} {C83}},\ \bibinfo {pages}
  {054911} (\bibinfo {year} {2011})},\ \Eprint {http://arxiv.org/abs/1103.4239}
  {arXiv:1103.4239 [nucl-th]} \BibitemShut {NoStop}%
\bibitem [{\citenamefont {Deng}\ and\ \citenamefont
  {Huang}(2012)}]{Deng:2012pc}%
  \BibitemOpen
  \bibfield  {author} {\bibinfo {author} {\bibfnamefont {W.-T.}\ \bibnamefont
  {Deng}}\ and\ \bibinfo {author} {\bibfnamefont {X.-G.}\ \bibnamefont
  {Huang}},\ }\href {\doibase 10.1103/PhysRevC.85.044907} {\bibfield  {journal}
  {\bibinfo  {journal} {Phys. Rev.}\ }\textbf {\bibinfo {volume} {C85}},\
  \bibinfo {pages} {044907} (\bibinfo {year} {2012})},\ \Eprint
  {http://arxiv.org/abs/1201.5108} {arXiv:1201.5108 [nucl-th]} \BibitemShut
  {NoStop}%
\bibitem [{\citenamefont {Brambilla}\ \emph {et~al.}(2014)\citenamefont
  {Brambilla} \emph {et~al.}}]{Brambilla:2014jmp}%
  \BibitemOpen
  \bibfield  {author} {\bibinfo {author} {\bibfnamefont {N.}~\bibnamefont
  {Brambilla}} \emph {et~al.},\ }\href {\doibase
  10.1140/epjc/s10052-014-2981-5} {\bibfield  {journal} {\bibinfo  {journal}
  {Eur. Phys. J.}\ }\textbf {\bibinfo {volume} {C74}},\ \bibinfo {pages} {2981}
  (\bibinfo {year} {2014})},\ \Eprint {http://arxiv.org/abs/1404.3723}
  {arXiv:1404.3723 [hep-ph]} \BibitemShut {NoStop}%
\bibitem [{\citenamefont {Kharzeev}\ and\ \citenamefont
  {Zhitnitsky}(2007)}]{Kharzeev:2007tn}%
  \BibitemOpen
  \bibfield  {author} {\bibinfo {author} {\bibfnamefont {D.}~\bibnamefont
  {Kharzeev}}\ and\ \bibinfo {author} {\bibfnamefont {A.}~\bibnamefont
  {Zhitnitsky}},\ }\href {\doibase 10.1016/j.nuclphysa.2007.10.001} {\bibfield
  {journal} {\bibinfo  {journal} {Nucl. Phys.}\ }\textbf {\bibinfo {volume}
  {A797}},\ \bibinfo {pages} {67} (\bibinfo {year} {2007})},\ \Eprint
  {http://arxiv.org/abs/0706.1026} {arXiv:0706.1026 [hep-ph]} \BibitemShut
  {NoStop}%
\bibitem [{\citenamefont {Kharzeev}\ \emph {et~al.}(2008)\citenamefont
  {Kharzeev}, \citenamefont {McLerran},\ and\ \citenamefont
  {Warringa}}]{Kharzeev:2007jp}%
  \BibitemOpen
  \bibfield  {author} {\bibinfo {author} {\bibfnamefont {D.~E.}\ \bibnamefont
  {Kharzeev}}, \bibinfo {author} {\bibfnamefont {L.~D.}\ \bibnamefont
  {McLerran}}, \ and\ \bibinfo {author} {\bibfnamefont {H.~J.}\ \bibnamefont
  {Warringa}},\ }\href {\doibase 10.1016/j.nuclphysa.2008.02.298} {\bibfield
  {journal} {\bibinfo  {journal} {Nucl. Phys.}\ }\textbf {\bibinfo {volume}
  {A803}},\ \bibinfo {pages} {227} (\bibinfo {year} {2008})},\ \Eprint
  {http://arxiv.org/abs/0711.0950} {arXiv:0711.0950 [hep-ph]} \BibitemShut
  {NoStop}%
\bibitem [{\citenamefont {Fukushima}\ \emph {et~al.}(2008)\citenamefont
  {Fukushima}, \citenamefont {Kharzeev},\ and\ \citenamefont
  {Warringa}}]{Fukushima:2008xe}%
  \BibitemOpen
  \bibfield  {author} {\bibinfo {author} {\bibfnamefont {K.}~\bibnamefont
  {Fukushima}}, \bibinfo {author} {\bibfnamefont {D.~E.}\ \bibnamefont
  {Kharzeev}}, \ and\ \bibinfo {author} {\bibfnamefont {H.~J.}\ \bibnamefont
  {Warringa}},\ }\href {\doibase 10.1103/PhysRevD.78.074033} {\bibfield
  {journal} {\bibinfo  {journal} {Phys. Rev.}\ }\textbf {\bibinfo {volume}
  {D78}},\ \bibinfo {pages} {074033} (\bibinfo {year} {2008})},\ \Eprint
  {http://arxiv.org/abs/0808.3382} {arXiv:0808.3382 [hep-ph]} \BibitemShut
  {NoStop}%
\bibitem [{\citenamefont {Son}\ and\ \citenamefont
  {Zhitnitsky}(2004)}]{Son:2004tq}%
  \BibitemOpen
  \bibfield  {author} {\bibinfo {author} {\bibfnamefont {D.~T.}\ \bibnamefont
  {Son}}\ and\ \bibinfo {author} {\bibfnamefont {A.~R.}\ \bibnamefont
  {Zhitnitsky}},\ }\href {\doibase 10.1103/PhysRevD.70.074018} {\bibfield
  {journal} {\bibinfo  {journal} {Phys. Rev.}\ }\textbf {\bibinfo {volume}
  {D70}},\ \bibinfo {pages} {074018} (\bibinfo {year} {2004})},\ \Eprint
  {http://arxiv.org/abs/hep-ph/0405216} {arXiv:hep-ph/0405216 [hep-ph]}
  \BibitemShut {NoStop}%
\bibitem [{\citenamefont {Metlitski}\ and\ \citenamefont
  {Zhitnitsky}(2005)}]{Metlitski:2005pr}%
  \BibitemOpen
  \bibfield  {author} {\bibinfo {author} {\bibfnamefont {M.~A.}\ \bibnamefont
  {Metlitski}}\ and\ \bibinfo {author} {\bibfnamefont {A.~R.}\ \bibnamefont
  {Zhitnitsky}},\ }\href {\doibase 10.1103/PhysRevD.72.045011} {\bibfield
  {journal} {\bibinfo  {journal} {Phys. Rev.}\ }\textbf {\bibinfo {volume}
  {D72}},\ \bibinfo {pages} {045011} (\bibinfo {year} {2005})},\ \Eprint
  {http://arxiv.org/abs/hep-ph/0505072} {arXiv:hep-ph/0505072 [hep-ph]}
  \BibitemShut {NoStop}%
\bibitem [{\citenamefont {Son}\ and\ \citenamefont
  {Surowka}(2009)}]{Son:2009tf}%
  \BibitemOpen
  \bibfield  {author} {\bibinfo {author} {\bibfnamefont {D.~T.}\ \bibnamefont
  {Son}}\ and\ \bibinfo {author} {\bibfnamefont {P.}~\bibnamefont {Surowka}},\
  }\href {\doibase 10.1103/PhysRevLett.103.191601} {\bibfield  {journal}
  {\bibinfo  {journal} {Phys. Rev. Lett.}\ }\textbf {\bibinfo {volume} {103}},\
  \bibinfo {pages} {191601} (\bibinfo {year} {2009})},\ \Eprint
  {http://arxiv.org/abs/0906.5044} {arXiv:0906.5044 [hep-th]} \BibitemShut
  {NoStop}%
\bibitem [{\citenamefont {Kharzeev}\ and\ \citenamefont
  {Yee}(2011)}]{Kharzeev:2010gd}%
  \BibitemOpen
  \bibfield  {author} {\bibinfo {author} {\bibfnamefont {D.~E.}\ \bibnamefont
  {Kharzeev}}\ and\ \bibinfo {author} {\bibfnamefont {H.-U.}\ \bibnamefont
  {Yee}},\ }\href {\doibase 10.1103/PhysRevD.83.085007} {\bibfield  {journal}
  {\bibinfo  {journal} {Phys. Rev.}\ }\textbf {\bibinfo {volume} {D83}},\
  \bibinfo {pages} {085007} (\bibinfo {year} {2011})},\ \Eprint
  {http://arxiv.org/abs/1012.6026} {arXiv:1012.6026 [hep-th]} \BibitemShut
  {NoStop}%
\bibitem [{\citenamefont {Fukushima}(2013)}]{Fukushima:2012vr}%
  \BibitemOpen
  \bibfield  {author} {\bibinfo {author} {\bibfnamefont {K.}~\bibnamefont
  {Fukushima}},\ }\href {\doibase 10.1007/978-3-642-37305-3_9} {\bibfield
  {journal} {\bibinfo  {journal} {Lect. Notes Phys.}\ }\textbf {\bibinfo
  {volume} {871}},\ \bibinfo {pages} {241} (\bibinfo {year} {2013})},\ \Eprint
  {http://arxiv.org/abs/1209.5064} {arXiv:1209.5064 [hep-ph]} \BibitemShut
  {NoStop}%
\bibitem [{\citenamefont {Kharzeev}(2014)}]{Kharzeev:2013ffa}%
  \BibitemOpen
  \bibfield  {author} {\bibinfo {author} {\bibfnamefont {D.~E.}\ \bibnamefont
  {Kharzeev}},\ }\href {\doibase 10.1016/j.ppnp.2014.01.002} {\bibfield
  {journal} {\bibinfo  {journal} {Prog. Part. Nucl. Phys.}\ }\textbf {\bibinfo
  {volume} {75}},\ \bibinfo {pages} {133} (\bibinfo {year} {2014})},\ \Eprint
  {http://arxiv.org/abs/1312.3348} {arXiv:1312.3348 [hep-ph]} \BibitemShut
  {NoStop}%
\bibitem [{\citenamefont {Huang}(2015{\natexlab{a}})}]{Huang:2015oca}%
  \BibitemOpen
  \bibfield  {author} {\bibinfo {author} {\bibfnamefont {X.-G.}\ \bibnamefont
  {Huang}},\ }\href@noop {} {\  (\bibinfo {year} {2015}{\natexlab{a}})},\
  \Eprint {http://arxiv.org/abs/1509.04073} {arXiv:1509.04073 [nucl-th]}
  \BibitemShut {NoStop}%
\bibitem [{\citenamefont {Kharzeev}\ \emph {et~al.}(2015)\citenamefont
  {Kharzeev}, \citenamefont {Liao}, \citenamefont {Voloshin},\ and\
  \citenamefont {Wang}}]{Kharzeev:2015znc}%
  \BibitemOpen
  \bibfield  {author} {\bibinfo {author} {\bibfnamefont {D.~E.}\ \bibnamefont
  {Kharzeev}}, \bibinfo {author} {\bibfnamefont {J.}~\bibnamefont {Liao}},
  \bibinfo {author} {\bibfnamefont {S.~A.}\ \bibnamefont {Voloshin}}, \ and\
  \bibinfo {author} {\bibfnamefont {G.}~\bibnamefont {Wang}},\ }\href@noop {}
  {\  (\bibinfo {year} {2015})},\ \Eprint {http://arxiv.org/abs/1511.04050}
  {arXiv:1511.04050 [hep-ph]} \BibitemShut {NoStop}%
\bibitem [{\citenamefont {Klimenko}(1992{\natexlab{a}})}]{Klimenko:1991he}%
  \BibitemOpen
  \bibfield  {author} {\bibinfo {author} {\bibfnamefont {K.~G.}\ \bibnamefont
  {Klimenko}},\ }\href {\doibase 10.1007/BF01566663} {\bibfield  {journal}
  {\bibinfo  {journal} {Z. Phys.}\ }\textbf {\bibinfo {volume} {C54}},\
  \bibinfo {pages} {323} (\bibinfo {year} {1992}{\natexlab{a}})}\BibitemShut
  {NoStop}%
\bibitem [{\citenamefont {Klimenko}(1992{\natexlab{b}})}]{Klimenko:1992ch}%
  \BibitemOpen
  \bibfield  {author} {\bibinfo {author} {\bibfnamefont {K.~G.}\ \bibnamefont
  {Klimenko}},\ }\href {\doibase 10.1007/BF01018812} {\bibfield  {journal}
  {\bibinfo  {journal} {Theor. Math. Phys.}\ }\textbf {\bibinfo {volume}
  {90}},\ \bibinfo {pages} {1} (\bibinfo {year} {1992}{\natexlab{b}})},\
  \bibinfo {note} {[Teor. Mat. Fiz.90,3(1992)]}\BibitemShut {NoStop}%
\bibitem [{\citenamefont {Gusynin}\ \emph {et~al.}(1994)\citenamefont
  {Gusynin}, \citenamefont {Miransky},\ and\ \citenamefont
  {Shovkovy}}]{Gusynin:1994re}%
  \BibitemOpen
  \bibfield  {author} {\bibinfo {author} {\bibfnamefont {V.~P.}\ \bibnamefont
  {Gusynin}}, \bibinfo {author} {\bibfnamefont {V.~A.}\ \bibnamefont
  {Miransky}}, \ and\ \bibinfo {author} {\bibfnamefont {I.~A.}\ \bibnamefont
  {Shovkovy}},\ }\href {\doibase 10.1103/PhysRevLett.73.3499} {\bibfield
  {journal} {\bibinfo  {journal} {Phys. Rev. Lett.}\ }\textbf {\bibinfo
  {volume} {73}},\ \bibinfo {pages} {3499} (\bibinfo {year} {1994})},\ \bibinfo
  {note} {[Erratum: Phys. Rev. Lett.76,1005(1996)]},\ \Eprint
  {http://arxiv.org/abs/hep-ph/9405262} {arXiv:hep-ph/9405262 [hep-ph]}
  \BibitemShut {NoStop}%
\bibitem [{\citenamefont {Gusynin}\ \emph {et~al.}(1996)\citenamefont
  {Gusynin}, \citenamefont {Miransky},\ and\ \citenamefont
  {Shovkovy}}]{Gusynin:1995nb}%
  \BibitemOpen
  \bibfield  {author} {\bibinfo {author} {\bibfnamefont {V.~P.}\ \bibnamefont
  {Gusynin}}, \bibinfo {author} {\bibfnamefont {V.~A.}\ \bibnamefont
  {Miransky}}, \ and\ \bibinfo {author} {\bibfnamefont {I.~A.}\ \bibnamefont
  {Shovkovy}},\ }\href {\doibase 10.1016/0550-3213(96)00021-1} {\bibfield
  {journal} {\bibinfo  {journal} {Nucl. Phys.}\ }\textbf {\bibinfo {volume}
  {B462}},\ \bibinfo {pages} {249} (\bibinfo {year} {1996})},\ \Eprint
  {http://arxiv.org/abs/hep-ph/9509320} {arXiv:hep-ph/9509320 [hep-ph]}
  \BibitemShut {NoStop}%
\bibitem [{\citenamefont {Shovkovy}(2013)}]{Shovkovy:2012zn}%
  \BibitemOpen
  \bibfield  {author} {\bibinfo {author} {\bibfnamefont {I.~A.}\ \bibnamefont
  {Shovkovy}},\ }\href {\doibase 10.1007/978-3-642-37305-3_2} {\bibfield
  {journal} {\bibinfo  {journal} {Lect. Notes Phys.}\ }\textbf {\bibinfo
  {volume} {871}},\ \bibinfo {pages} {13} (\bibinfo {year} {2013})},\ \Eprint
  {http://arxiv.org/abs/1207.5081} {arXiv:1207.5081 [hep-ph]} \BibitemShut
  {NoStop}%
\bibitem [{\citenamefont {Miransky}\ and\ \citenamefont
  {Shovkovy}(2015)}]{Miransky:2015ava}%
  \BibitemOpen
  \bibfield  {author} {\bibinfo {author} {\bibfnamefont {V.~A.}\ \bibnamefont
  {Miransky}}\ and\ \bibinfo {author} {\bibfnamefont {I.~A.}\ \bibnamefont
  {Shovkovy}},\ }\href {\doibase 10.1016/j.physrep.2015.02.003} {\bibfield
  {journal} {\bibinfo  {journal} {Phys. Rept.}\ }\textbf {\bibinfo {volume}
  {576}},\ \bibinfo {pages} {1} (\bibinfo {year} {2015})},\ \Eprint
  {http://arxiv.org/abs/1503.00732} {arXiv:1503.00732 [hep-ph]} \BibitemShut
  {NoStop}%
\bibitem [{\citenamefont {Ebert}\ \emph {et~al.}(1999)\citenamefont {Ebert},
  \citenamefont {Klimenko}, \citenamefont {Vdovichenko},\ and\ \citenamefont
  {Vshivtsev}}]{Ebert:1999ht}%
  \BibitemOpen
  \bibfield  {author} {\bibinfo {author} {\bibfnamefont {D.}~\bibnamefont
  {Ebert}}, \bibinfo {author} {\bibfnamefont {K.~G.}\ \bibnamefont {Klimenko}},
  \bibinfo {author} {\bibfnamefont {M.~A.}\ \bibnamefont {Vdovichenko}}, \ and\
  \bibinfo {author} {\bibfnamefont {A.~S.}\ \bibnamefont {Vshivtsev}},\ }\href
  {\doibase 10.1103/PhysRevD.61.025005} {\bibfield  {journal} {\bibinfo
  {journal} {Phys. Rev.}\ }\textbf {\bibinfo {volume} {D61}},\ \bibinfo {pages}
  {025005} (\bibinfo {year} {1999})},\ \Eprint
  {http://arxiv.org/abs/hep-ph/9905253} {arXiv:hep-ph/9905253 [hep-ph]}
  \BibitemShut {NoStop}%
\bibitem [{\citenamefont {Inagaki}\ \emph {et~al.}(2004)\citenamefont
  {Inagaki}, \citenamefont {Kimura},\ and\ \citenamefont
  {Murata}}]{Inagaki:2003yi}%
  \BibitemOpen
  \bibfield  {author} {\bibinfo {author} {\bibfnamefont {T.}~\bibnamefont
  {Inagaki}}, \bibinfo {author} {\bibfnamefont {D.}~\bibnamefont {Kimura}}, \
  and\ \bibinfo {author} {\bibfnamefont {T.}~\bibnamefont {Murata}},\ }\href
  {\doibase 10.1143/PTP.111.371} {\bibfield  {journal} {\bibinfo  {journal}
  {Prog. Theor. Phys.}\ }\textbf {\bibinfo {volume} {111}},\ \bibinfo {pages}
  {371} (\bibinfo {year} {2004})},\ \Eprint
  {http://arxiv.org/abs/hep-ph/0312005} {arXiv:hep-ph/0312005 [hep-ph]}
  \BibitemShut {NoStop}%
\bibitem [{\citenamefont {Fraga}\ and\ \citenamefont
  {Mizher}(2008)}]{Fraga:2008qn}%
  \BibitemOpen
  \bibfield  {author} {\bibinfo {author} {\bibfnamefont {E.~S.}\ \bibnamefont
  {Fraga}}\ and\ \bibinfo {author} {\bibfnamefont {A.~J.}\ \bibnamefont
  {Mizher}},\ }\href {\doibase 10.1103/PhysRevD.78.025016} {\bibfield
  {journal} {\bibinfo  {journal} {Phys. Rev.}\ }\textbf {\bibinfo {volume}
  {D78}},\ \bibinfo {pages} {025016} (\bibinfo {year} {2008})},\ \Eprint
  {http://arxiv.org/abs/0804.1452} {arXiv:0804.1452 [hep-ph]} \BibitemShut
  {NoStop}%
\bibitem [{\citenamefont {Mizher}\ \emph {et~al.}(2010)\citenamefont {Mizher},
  \citenamefont {Chernodub},\ and\ \citenamefont {Fraga}}]{Mizher:2010zb}%
  \BibitemOpen
  \bibfield  {author} {\bibinfo {author} {\bibfnamefont {A.~J.}\ \bibnamefont
  {Mizher}}, \bibinfo {author} {\bibfnamefont {M.~N.}\ \bibnamefont
  {Chernodub}}, \ and\ \bibinfo {author} {\bibfnamefont {E.~S.}\ \bibnamefont
  {Fraga}},\ }\href {\doibase 10.1103/PhysRevD.82.105016} {\bibfield  {journal}
  {\bibinfo  {journal} {Phys. Rev.}\ }\textbf {\bibinfo {volume} {D82}},\
  \bibinfo {pages} {105016} (\bibinfo {year} {2010})},\ \Eprint
  {http://arxiv.org/abs/1004.2712} {arXiv:1004.2712 [hep-ph]} \BibitemShut
  {NoStop}%
\bibitem [{\citenamefont {Andersen}\ and\ \citenamefont
  {Khan}(2012)}]{Andersen:2011ip}%
  \BibitemOpen
  \bibfield  {author} {\bibinfo {author} {\bibfnamefont {J.~O.}\ \bibnamefont
  {Andersen}}\ and\ \bibinfo {author} {\bibfnamefont {R.}~\bibnamefont
  {Khan}},\ }\href {\doibase 10.1103/PhysRevD.85.065026} {\bibfield  {journal}
  {\bibinfo  {journal} {Phys. Rev.}\ }\textbf {\bibinfo {volume} {D85}},\
  \bibinfo {pages} {065026} (\bibinfo {year} {2012})},\ \Eprint
  {http://arxiv.org/abs/1105.1290} {arXiv:1105.1290 [hep-ph]} \BibitemShut
  {NoStop}%
\bibitem [{\citenamefont {Ferrari}\ \emph {et~al.}(2012)\citenamefont
  {Ferrari}, \citenamefont {Garcia},\ and\ \citenamefont
  {Pinto}}]{Ferrari:2012yw}%
  \BibitemOpen
  \bibfield  {author} {\bibinfo {author} {\bibfnamefont {G.~N.}\ \bibnamefont
  {Ferrari}}, \bibinfo {author} {\bibfnamefont {A.~F.}\ \bibnamefont {Garcia}},
  \ and\ \bibinfo {author} {\bibfnamefont {M.~B.}\ \bibnamefont {Pinto}},\
  }\href {\doibase 10.1103/PhysRevD.86.096005} {\bibfield  {journal} {\bibinfo
  {journal} {Phys. Rev.}\ }\textbf {\bibinfo {volume} {D86}},\ \bibinfo {pages}
  {096005} (\bibinfo {year} {2012})},\ \Eprint {http://arxiv.org/abs/1207.3714}
  {arXiv:1207.3714 [hep-ph]} \BibitemShut {NoStop}%
\bibitem [{\citenamefont {Fraga}\ and\ \citenamefont
  {Palhares}(2012)}]{Fraga:2012fs}%
  \BibitemOpen
  \bibfield  {author} {\bibinfo {author} {\bibfnamefont {E.~S.}\ \bibnamefont
  {Fraga}}\ and\ \bibinfo {author} {\bibfnamefont {L.~F.}\ \bibnamefont
  {Palhares}},\ }\href {\doibase 10.1103/PhysRevD.86.016008} {\bibfield
  {journal} {\bibinfo  {journal} {Phys. Rev.}\ }\textbf {\bibinfo {volume}
  {D86}},\ \bibinfo {pages} {016008} (\bibinfo {year} {2012})},\ \Eprint
  {http://arxiv.org/abs/1201.5881} {arXiv:1201.5881 [hep-ph]} \BibitemShut
  {NoStop}%
\bibitem [{\citenamefont {Bali}\ \emph
  {et~al.}(2012{\natexlab{a}})\citenamefont {Bali}, \citenamefont {Bruckmann},
  \citenamefont {Endrodi}, \citenamefont {Fodor}, \citenamefont {Katz},\ and\
  \citenamefont {Schafer}}]{Bali:2012zg}%
  \BibitemOpen
  \bibfield  {author} {\bibinfo {author} {\bibfnamefont {G.~S.}\ \bibnamefont
  {Bali}}, \bibinfo {author} {\bibfnamefont {F.}~\bibnamefont {Bruckmann}},
  \bibinfo {author} {\bibfnamefont {G.}~\bibnamefont {Endrodi}}, \bibinfo
  {author} {\bibfnamefont {Z.}~\bibnamefont {Fodor}}, \bibinfo {author}
  {\bibfnamefont {S.~D.}\ \bibnamefont {Katz}}, \ and\ \bibinfo {author}
  {\bibfnamefont {A.}~\bibnamefont {Schafer}},\ }\href {\doibase
  10.1103/PhysRevD.86.071502} {\bibfield  {journal} {\bibinfo  {journal} {Phys.
  Rev.}\ }\textbf {\bibinfo {volume} {D86}},\ \bibinfo {pages} {071502}
  (\bibinfo {year} {2012}{\natexlab{a}})},\ \Eprint
  {http://arxiv.org/abs/1206.4205} {arXiv:1206.4205 [hep-lat]} \BibitemShut
  {NoStop}%
\bibitem [{\citenamefont {Johnson}\ and\ \citenamefont
  {Kundu}(2008)}]{Johnson:2008vna}%
  \BibitemOpen
  \bibfield  {author} {\bibinfo {author} {\bibfnamefont {C.~V.}\ \bibnamefont
  {Johnson}}\ and\ \bibinfo {author} {\bibfnamefont {A.}~\bibnamefont
  {Kundu}},\ }\href {\doibase 10.1088/1126-6708/2008/12/053} {\bibfield
  {journal} {\bibinfo  {journal} {JHEP}\ }\textbf {\bibinfo {volume} {12}},\
  \bibinfo {pages} {053} (\bibinfo {year} {2008})},\ \Eprint
  {http://arxiv.org/abs/0803.0038} {arXiv:0803.0038 [hep-th]} \BibitemShut
  {NoStop}%
\bibitem [{\citenamefont {Gorbar}\ \emph
  {et~al.}(2008{\natexlab{a}})\citenamefont {Gorbar}, \citenamefont {Gusynin},\
  and\ \citenamefont {Miransky}}]{Gorbar:2007xh}%
  \BibitemOpen
  \bibfield  {author} {\bibinfo {author} {\bibfnamefont {E.~V.}\ \bibnamefont
  {Gorbar}}, \bibinfo {author} {\bibfnamefont {V.~P.}\ \bibnamefont {Gusynin}},
  \ and\ \bibinfo {author} {\bibfnamefont {V.~A.}\ \bibnamefont {Miransky}},\
  }\href {\doibase 10.1063/1.2981388} {\bibfield  {journal} {\bibinfo
  {journal} {Low Temp. Phys.}\ }\textbf {\bibinfo {volume} {34}},\ \bibinfo
  {pages} {790} (\bibinfo {year} {2008}{\natexlab{a}})},\ \Eprint
  {http://arxiv.org/abs/0710.3527} {arXiv:0710.3527 [cond-mat.mes-hall]}
  \BibitemShut {NoStop}%
\bibitem [{\citenamefont {Gorbar}\ \emph
  {et~al.}(2008{\natexlab{b}})\citenamefont {Gorbar}, \citenamefont {Gusynin},
  \citenamefont {Miransky},\ and\ \citenamefont {Shovkovy}}]{Gorbar:2008hu}%
  \BibitemOpen
  \bibfield  {author} {\bibinfo {author} {\bibfnamefont {E.~V.}\ \bibnamefont
  {Gorbar}}, \bibinfo {author} {\bibfnamefont {V.~P.}\ \bibnamefont {Gusynin}},
  \bibinfo {author} {\bibfnamefont {V.~A.}\ \bibnamefont {Miransky}}, \ and\
  \bibinfo {author} {\bibfnamefont {I.~A.}\ \bibnamefont {Shovkovy}},\ }\href
  {\doibase 10.1103/PhysRevB.78.085437} {\bibfield  {journal} {\bibinfo
  {journal} {Phys. Rev.}\ }\textbf {\bibinfo {volume} {B78}},\ \bibinfo {pages}
  {085437} (\bibinfo {year} {2008}{\natexlab{b}})},\ \Eprint
  {http://arxiv.org/abs/0806.0846} {arXiv:0806.0846 [cond-mat.mes-hall]}
  \BibitemShut {NoStop}%
\bibitem [{\citenamefont {Roy}\ and\ \citenamefont {Sau}(2015)}]{Roy:2014mia}%
  \BibitemOpen
  \bibfield  {author} {\bibinfo {author} {\bibfnamefont {B.}~\bibnamefont
  {Roy}}\ and\ \bibinfo {author} {\bibfnamefont {J.~D.}\ \bibnamefont {Sau}},\
  }\href {\doibase 10.1103/PhysRevB.92.125141} {\bibfield  {journal} {\bibinfo
  {journal} {Phys. Rev.}\ }\textbf {\bibinfo {volume} {B92}},\ \bibinfo {pages}
  {125141} (\bibinfo {year} {2015})},\ \Eprint {http://arxiv.org/abs/1406.4501}
  {arXiv:1406.4501 [cond-mat.mes-hall]} \BibitemShut {NoStop}%
\bibitem [{\citenamefont {Ayala}\ \emph {et~al.}(2012)\citenamefont {Ayala},
  \citenamefont {Loewe}, \citenamefont {Rojas},\ and\ \citenamefont
  {Villavicencio}}]{Ayala:2012dk}%
  \BibitemOpen
  \bibfield  {author} {\bibinfo {author} {\bibfnamefont {A.}~\bibnamefont
  {Ayala}}, \bibinfo {author} {\bibfnamefont {M.}~\bibnamefont {Loewe}},
  \bibinfo {author} {\bibfnamefont {J.~C.}\ \bibnamefont {Rojas}}, \ and\
  \bibinfo {author} {\bibfnamefont {C.}~\bibnamefont {Villavicencio}},\ }\href
  {\doibase 10.1103/PhysRevD.86.076006, 10.1103/PhysRevD.86.079902} {\bibfield
  {journal} {\bibinfo  {journal} {Phys. Rev.}\ }\textbf {\bibinfo {volume}
  {D86}},\ \bibinfo {pages} {076006} (\bibinfo {year} {2012})},\ \Eprint
  {http://arxiv.org/abs/1208.0390} {arXiv:1208.0390 [hep-ph]} \BibitemShut
  {NoStop}%
\bibitem [{\citenamefont {Feng}\ \emph {et~al.}(2015)\citenamefont {Feng},
  \citenamefont {Hou},\ and\ \citenamefont {Ren}}]{Feng:2014bpa}%
  \BibitemOpen
  \bibfield  {author} {\bibinfo {author} {\bibfnamefont {B.}~\bibnamefont
  {Feng}}, \bibinfo {author} {\bibfnamefont {D.-F.}\ \bibnamefont {Hou}}, \
  and\ \bibinfo {author} {\bibfnamefont {H.-C.}\ \bibnamefont {Ren}},\ }\href
  {\doibase 10.1103/PhysRevD.92.065011} {\bibfield  {journal} {\bibinfo
  {journal} {Phys. Rev.}\ }\textbf {\bibinfo {volume} {D92}},\ \bibinfo {pages}
  {065011} (\bibinfo {year} {2015})},\ \Eprint {http://arxiv.org/abs/1412.1647}
  {arXiv:1412.1647 [cond-mat.quant-gas]} \BibitemShut {NoStop}%
\bibitem [{\citenamefont {Preis}\ \emph {et~al.}(2011)\citenamefont {Preis},
  \citenamefont {Rebhan},\ and\ \citenamefont {Schmitt}}]{Preis:2010cq}%
  \BibitemOpen
  \bibfield  {author} {\bibinfo {author} {\bibfnamefont {F.}~\bibnamefont
  {Preis}}, \bibinfo {author} {\bibfnamefont {A.}~\bibnamefont {Rebhan}}, \
  and\ \bibinfo {author} {\bibfnamefont {A.}~\bibnamefont {Schmitt}},\ }\href
  {\doibase 10.1007/JHEP03(2011)033} {\bibfield  {journal} {\bibinfo  {journal}
  {JHEP}\ }\textbf {\bibinfo {volume} {03}},\ \bibinfo {pages} {033} (\bibinfo
  {year} {2011})},\ \Eprint {http://arxiv.org/abs/1012.4785} {arXiv:1012.4785
  [hep-th]} \BibitemShut {NoStop}%
\bibitem [{\citenamefont {Fukushima}\ and\ \citenamefont
  {Hidaka}(2013)}]{Fukushima:2012kc}%
  \BibitemOpen
  \bibfield  {author} {\bibinfo {author} {\bibfnamefont {K.}~\bibnamefont
  {Fukushima}}\ and\ \bibinfo {author} {\bibfnamefont {Y.}~\bibnamefont
  {Hidaka}},\ }\href {\doibase 10.1103/PhysRevLett.110.031601} {\bibfield
  {journal} {\bibinfo  {journal} {Phys. Rev. Lett.}\ }\textbf {\bibinfo
  {volume} {110}},\ \bibinfo {pages} {031601} (\bibinfo {year} {2013})},\
  \Eprint {http://arxiv.org/abs/1209.1319} {arXiv:1209.1319 [hep-ph]}
  \BibitemShut {NoStop}%
\bibitem [{\citenamefont {Bali}\ \emph
  {et~al.}(2012{\natexlab{b}})\citenamefont {Bali}, \citenamefont {Bruckmann},
  \citenamefont {Endrodi}, \citenamefont {Fodor}, \citenamefont {Katz},
  \citenamefont {Krieg}, \citenamefont {Schafer},\ and\ \citenamefont
  {Szabo}}]{Bali:2011qj}%
  \BibitemOpen
  \bibfield  {author} {\bibinfo {author} {\bibfnamefont {G.~S.}\ \bibnamefont
  {Bali}}, \bibinfo {author} {\bibfnamefont {F.}~\bibnamefont {Bruckmann}},
  \bibinfo {author} {\bibfnamefont {G.}~\bibnamefont {Endrodi}}, \bibinfo
  {author} {\bibfnamefont {Z.}~\bibnamefont {Fodor}}, \bibinfo {author}
  {\bibfnamefont {S.~D.}\ \bibnamefont {Katz}}, \bibinfo {author}
  {\bibfnamefont {S.}~\bibnamefont {Krieg}}, \bibinfo {author} {\bibfnamefont
  {A.}~\bibnamefont {Schafer}}, \ and\ \bibinfo {author} {\bibfnamefont
  {K.~K.}\ \bibnamefont {Szabo}},\ }\href {\doibase 10.1007/JHEP02(2012)044}
  {\bibfield  {journal} {\bibinfo  {journal} {JHEP}\ }\textbf {\bibinfo
  {volume} {02}},\ \bibinfo {pages} {044} (\bibinfo {year}
  {2012}{\natexlab{b}})},\ \Eprint {http://arxiv.org/abs/1111.4956}
  {arXiv:1111.4956 [hep-lat]} \BibitemShut {NoStop}%
\bibitem [{\citenamefont {Fukushima}\ and\ \citenamefont
  {Pawlowski}(2012)}]{Fukushima:2012xw}%
  \BibitemOpen
  \bibfield  {author} {\bibinfo {author} {\bibfnamefont {K.}~\bibnamefont
  {Fukushima}}\ and\ \bibinfo {author} {\bibfnamefont {J.~M.}\ \bibnamefont
  {Pawlowski}},\ }\href {\doibase 10.1103/PhysRevD.86.076013} {\bibfield
  {journal} {\bibinfo  {journal} {Phys. Rev.}\ }\textbf {\bibinfo {volume}
  {D86}},\ \bibinfo {pages} {076013} (\bibinfo {year} {2012})},\ \Eprint
  {http://arxiv.org/abs/1203.4330} {arXiv:1203.4330 [hep-ph]} \BibitemShut
  {NoStop}%
\bibitem [{\citenamefont {Preis}\ \emph {et~al.}(2013)\citenamefont {Preis},
  \citenamefont {Rebhan},\ and\ \citenamefont {Schmitt}}]{Preis:2012fh}%
  \BibitemOpen
  \bibfield  {author} {\bibinfo {author} {\bibfnamefont {F.}~\bibnamefont
  {Preis}}, \bibinfo {author} {\bibfnamefont {A.}~\bibnamefont {Rebhan}}, \
  and\ \bibinfo {author} {\bibfnamefont {A.}~\bibnamefont {Schmitt}},\ }\href
  {\doibase 10.1007/978-3-642-37305-3_3} {\bibfield  {journal} {\bibinfo
  {journal} {Lect. Notes Phys.}\ }\textbf {\bibinfo {volume} {871}},\ \bibinfo
  {pages} {51} (\bibinfo {year} {2013})},\ \Eprint
  {http://arxiv.org/abs/1208.0536} {arXiv:1208.0536 [hep-ph]} \BibitemShut
  {NoStop}%
\bibitem [{\citenamefont {Bruckmann}\ \emph {et~al.}(2013)\citenamefont
  {Bruckmann}, \citenamefont {Endrodi},\ and\ \citenamefont
  {Kovacs}}]{Bruckmann:2013oba}%
  \BibitemOpen
  \bibfield  {author} {\bibinfo {author} {\bibfnamefont {F.}~\bibnamefont
  {Bruckmann}}, \bibinfo {author} {\bibfnamefont {G.}~\bibnamefont {Endrodi}},
  \ and\ \bibinfo {author} {\bibfnamefont {T.~G.}\ \bibnamefont {Kovacs}},\
  }\href {\doibase 10.1007/JHEP04(2013)112} {\bibfield  {journal} {\bibinfo
  {journal} {JHEP}\ }\textbf {\bibinfo {volume} {04}},\ \bibinfo {pages} {112}
  (\bibinfo {year} {2013})},\ \Eprint {http://arxiv.org/abs/1303.3972}
  {arXiv:1303.3972 [hep-lat]} \BibitemShut {NoStop}%
\bibitem [{\citenamefont {Wilkin}\ and\ \citenamefont
  {Gunn}(2000)}]{Wilkin:2000zz}%
  \BibitemOpen
  \bibfield  {author} {\bibinfo {author} {\bibfnamefont {N.~K.}\ \bibnamefont
  {Wilkin}}\ and\ \bibinfo {author} {\bibfnamefont {J.~M.~F.}\ \bibnamefont
  {Gunn}},\ }\href {\doibase 10.1103/PhysRevLett.84.6} {\bibfield  {journal}
  {\bibinfo  {journal} {Phys. Rev. Lett.}\ }\textbf {\bibinfo {volume} {84}},\
  \bibinfo {pages} {6} (\bibinfo {year} {2000})}\BibitemShut {NoStop}%
\bibitem [{\citenamefont {Cooper}\ \emph {et~al.}(2001)\citenamefont {Cooper},
  \citenamefont {Wilkin},\ and\ \citenamefont {Gunn}}]{Cooper:2001zz}%
  \BibitemOpen
  \bibfield  {author} {\bibinfo {author} {\bibfnamefont {N.~R.}\ \bibnamefont
  {Cooper}}, \bibinfo {author} {\bibfnamefont {N.~K.}\ \bibnamefont {Wilkin}},
  \ and\ \bibinfo {author} {\bibfnamefont {J.~M.~F.}\ \bibnamefont {Gunn}},\
  }\href {\doibase 10.1103/PhysRevLett.87.120405} {\bibfield  {journal}
  {\bibinfo  {journal} {Phys. Rev. Lett.}\ }\textbf {\bibinfo {volume} {87}},\
  \bibinfo {pages} {120405} (\bibinfo {year} {2001})}\BibitemShut {NoStop}%
\bibitem [{\citenamefont {Schweikhard}\ \emph {et~al.}(2004)\citenamefont
  {Schweikhard}, \citenamefont {Coddington}, \citenamefont {Engels},
  \citenamefont {Mogendorff},\ and\ \citenamefont
  {Cornell}}]{Schweikhard:2004zz}%
  \BibitemOpen
  \bibfield  {author} {\bibinfo {author} {\bibfnamefont {V.}~\bibnamefont
  {Schweikhard}}, \bibinfo {author} {\bibfnamefont {I.}~\bibnamefont
  {Coddington}}, \bibinfo {author} {\bibfnamefont {P.}~\bibnamefont {Engels}},
  \bibinfo {author} {\bibfnamefont {V.~P.}\ \bibnamefont {Mogendorff}}, \ and\
  \bibinfo {author} {\bibfnamefont {E.~A.}\ \bibnamefont {Cornell}},\ }\href
  {\doibase 10.1103/PhysRevLett.92.040404} {\bibfield  {journal} {\bibinfo
  {journal} {Phys. Rev. Lett.}\ }\textbf {\bibinfo {volume} {92}},\ \bibinfo
  {pages} {040404} (\bibinfo {year} {2004})},\ \Eprint
  {http://arxiv.org/abs/cond-mat/0308582} {arXiv:cond-mat/0308582
  [cond-mat.mes-hall]} \BibitemShut {NoStop}%
\bibitem [{\citenamefont {Mameda}\ and\ \citenamefont
  {Yamamoto}(2015)}]{Mameda:2015ria}%
  \BibitemOpen
  \bibfield  {author} {\bibinfo {author} {\bibfnamefont {K.}~\bibnamefont
  {Mameda}}\ and\ \bibinfo {author} {\bibfnamefont {A.}~\bibnamefont
  {Yamamoto}},\ }\href@noop {} {\  (\bibinfo {year} {2015})},\ \Eprint
  {http://arxiv.org/abs/1504.05826} {arXiv:1504.05826 [hep-th]} \BibitemShut
  {NoStop}%
\bibitem [{\citenamefont {Viefers}(2008)}]{viefers2008quantum}%
  \BibitemOpen
  \bibfield  {author} {\bibinfo {author} {\bibfnamefont {S.}~\bibnamefont
  {Viefers}},\ }\href@noop {} {\bibfield  {journal} {\bibinfo  {journal} {J. of
  Phys.: Cond. Matt.}\ }\textbf {\bibinfo {volume} {20}},\ \bibinfo {pages}
  {123202} (\bibinfo {year} {2008})}\BibitemShut {NoStop}%
\bibitem [{\citenamefont {Tsubota}\ \emph {et~al.}(2013)\citenamefont
  {Tsubota}, \citenamefont {Kasamatsu},\ and\ \citenamefont
  {Kobayashi}}]{tsubota2013novel}%
  \BibitemOpen
  \bibfield  {author} {\bibinfo {author} {\bibfnamefont {M.}~\bibnamefont
  {Tsubota}}, \bibinfo {author} {\bibfnamefont {K.}~\bibnamefont {Kasamatsu}},
  \ and\ \bibinfo {author} {\bibfnamefont {M.}~\bibnamefont {Kobayashi}},\
  }\href@noop {} {\emph {\bibinfo {title} {Novel Superfluids: Volume 1}}}\
  (\bibinfo  {publisher} {Oxford University Press},\ \bibinfo {year}
  {2013})\BibitemShut {NoStop}%
\bibitem [{\citenamefont {Huang}(2015{\natexlab{b}})}]{Huang:2015mga}%
  \BibitemOpen
  \bibfield  {author} {\bibinfo {author} {\bibfnamefont {X.-G.}\ \bibnamefont
  {Huang}},\ }\href@noop {} {\  (\bibinfo {year} {2015}{\natexlab{b}})},\
  \Eprint {http://arxiv.org/abs/1506.03590} {arXiv:1506.03590
  [cond-mat.quant-gas]} \BibitemShut {NoStop}%
\bibitem [{\citenamefont {Parker}\ and\ \citenamefont
  {Toms}(2009)}]{parker2009quantum}%
  \BibitemOpen
  \bibfield  {author} {\bibinfo {author} {\bibfnamefont {L.}~\bibnamefont
  {Parker}}\ and\ \bibinfo {author} {\bibfnamefont {D.}~\bibnamefont {Toms}},\
  }\href@noop {} {\emph {\bibinfo {title} {Quantum field theory in curved
  spacetime: quantized fields and gravity}}}\ (\bibinfo  {publisher} {Cambridge
  University Press},\ \bibinfo {year} {2009})\BibitemShut {NoStop}%
\bibitem [{\citenamefont {Becattini}\ and\ \citenamefont
  {Ferroni}(2007)}]{Becattini:2007zn}%
  \BibitemOpen
  \bibfield  {author} {\bibinfo {author} {\bibfnamefont {F.}~\bibnamefont
  {Becattini}}\ and\ \bibinfo {author} {\bibfnamefont {L.}~\bibnamefont
  {Ferroni}},\ }\href {\doibase 10.1140/epjc/s10052-007-0403-7} {\bibfield
  {journal} {\bibinfo  {journal} {Eur. Phys. J.}\ }\textbf {\bibinfo {volume}
  {C52}},\ \bibinfo {pages} {597} (\bibinfo {year} {2007})},\ \Eprint
  {http://arxiv.org/abs/0707.0793} {arXiv:0707.0793 [nucl-th]} \BibitemShut
  {NoStop}%
\bibitem [{\citenamefont {Becattini}\ and\ \citenamefont
  {Piccinini}(2008)}]{Becattini:2007nd}%
  \BibitemOpen
  \bibfield  {author} {\bibinfo {author} {\bibfnamefont {F.}~\bibnamefont
  {Becattini}}\ and\ \bibinfo {author} {\bibfnamefont {F.}~\bibnamefont
  {Piccinini}},\ }\href {\doibase 10.1016/j.aop.2008.01.001} {\bibfield
  {journal} {\bibinfo  {journal} {Annals Phys.}\ }\textbf {\bibinfo {volume}
  {323}},\ \bibinfo {pages} {2452} (\bibinfo {year} {2008})},\ \Eprint
  {http://arxiv.org/abs/0710.5694} {arXiv:0710.5694 [nucl-th]} \BibitemShut
  {NoStop}%
\bibitem [{\citenamefont {Becattini}\ and\ \citenamefont
  {Tinti}(2010)}]{Becattini:2009wh}%
  \BibitemOpen
  \bibfield  {author} {\bibinfo {author} {\bibfnamefont {F.}~\bibnamefont
  {Becattini}}\ and\ \bibinfo {author} {\bibfnamefont {L.}~\bibnamefont
  {Tinti}},\ }\href {\doibase 10.1016/j.aop.2010.03.007} {\bibfield  {journal}
  {\bibinfo  {journal} {Annals Phys.}\ }\textbf {\bibinfo {volume} {325}},\
  \bibinfo {pages} {1566} (\bibinfo {year} {2010})},\ \Eprint
  {http://arxiv.org/abs/0911.0864} {arXiv:0911.0864 [gr-qc]} \BibitemShut
  {NoStop}%
\bibitem [{\citenamefont {Becattini}\ \emph {et~al.}(2013)\citenamefont
  {Becattini}, \citenamefont {Chandra}, \citenamefont {Del~Zanna},\ and\
  \citenamefont {Grossi}}]{Becattini:2013fla}%
  \BibitemOpen
  \bibfield  {author} {\bibinfo {author} {\bibfnamefont {F.}~\bibnamefont
  {Becattini}}, \bibinfo {author} {\bibfnamefont {V.}~\bibnamefont {Chandra}},
  \bibinfo {author} {\bibfnamefont {L.}~\bibnamefont {Del~Zanna}}, \ and\
  \bibinfo {author} {\bibfnamefont {E.}~\bibnamefont {Grossi}},\ }\href
  {\doibase 10.1016/j.aop.2013.07.004} {\bibfield  {journal} {\bibinfo
  {journal} {Annals Phys.}\ }\textbf {\bibinfo {volume} {338}},\ \bibinfo
  {pages} {32} (\bibinfo {year} {2013})},\ \Eprint
  {http://arxiv.org/abs/1303.3431} {arXiv:1303.3431 [nucl-th]} \BibitemShut
  {NoStop}%
\bibitem [{\citenamefont {Elizalde}\ \emph {et~al.}(1994)\citenamefont
  {Elizalde}, \citenamefont {Leseduarte},\ and\ \citenamefont
  {Odintsov}}]{Elizalde:1993kb}%
  \BibitemOpen
  \bibfield  {author} {\bibinfo {author} {\bibfnamefont {E.}~\bibnamefont
  {Elizalde}}, \bibinfo {author} {\bibfnamefont {S.}~\bibnamefont
  {Leseduarte}}, \ and\ \bibinfo {author} {\bibfnamefont {S.~D.}\ \bibnamefont
  {Odintsov}},\ }\href {\doibase 10.1103/PhysRevD.49.5551} {\bibfield
  {journal} {\bibinfo  {journal} {Phys. Rev.}\ }\textbf {\bibinfo {volume}
  {D49}},\ \bibinfo {pages} {5551} (\bibinfo {year} {1994})},\ \Eprint
  {http://arxiv.org/abs/hep-th/9312164} {arXiv:hep-th/9312164 [hep-th]}
  \BibitemShut {NoStop}%
\bibitem [{\citenamefont {Inagaki}\ \emph {et~al.}(1993)\citenamefont
  {Inagaki}, \citenamefont {Muta},\ and\ \citenamefont
  {Odintsov}}]{Inagaki:1993ya}%
  \BibitemOpen
  \bibfield  {author} {\bibinfo {author} {\bibfnamefont {T.}~\bibnamefont
  {Inagaki}}, \bibinfo {author} {\bibfnamefont {T.}~\bibnamefont {Muta}}, \
  and\ \bibinfo {author} {\bibfnamefont {S.~D.}\ \bibnamefont {Odintsov}},\
  }\href {\doibase 10.1142/S0217732393001835} {\bibfield  {journal} {\bibinfo
  {journal} {Mod. Phys. Lett.}\ }\textbf {\bibinfo {volume} {A08}},\ \bibinfo
  {pages} {2117} (\bibinfo {year} {1993})},\ \Eprint
  {http://arxiv.org/abs/hep-th/9306023} {arXiv:hep-th/9306023 [hep-th]}
  \BibitemShut {NoStop}%
\bibitem [{\citenamefont {Inagaki}\ \emph {et~al.}(1997)\citenamefont
  {Inagaki}, \citenamefont {Muta},\ and\ \citenamefont
  {Odintsov}}]{Inagaki:1997kz}%
  \BibitemOpen
  \bibfield  {author} {\bibinfo {author} {\bibfnamefont {T.}~\bibnamefont
  {Inagaki}}, \bibinfo {author} {\bibfnamefont {T.}~\bibnamefont {Muta}}, \
  and\ \bibinfo {author} {\bibfnamefont {S.~D.}\ \bibnamefont {Odintsov}},\
  }\href {\doibase 10.1143/PTPS.127.93} {\bibfield  {journal} {\bibinfo
  {journal} {Prog. Theor. Phys. Suppl.}\ }\textbf {\bibinfo {volume} {127}},\
  \bibinfo {pages} {93} (\bibinfo {year} {1997})},\ \Eprint
  {http://arxiv.org/abs/hep-th/9711084} {arXiv:hep-th/9711084 [hep-th]}
  \BibitemShut {NoStop}%
\bibitem [{\citenamefont {Gorbar}(1999)}]{Gorbar:1999wa}%
  \BibitemOpen
  \bibfield  {author} {\bibinfo {author} {\bibfnamefont {E.~V.}\ \bibnamefont
  {Gorbar}},\ }\href {\doibase 10.1103/PhysRevD.61.024013} {\bibfield
  {journal} {\bibinfo  {journal} {Phys. Rev.}\ }\textbf {\bibinfo {volume}
  {D61}},\ \bibinfo {pages} {024013} (\bibinfo {year} {1999})},\ \Eprint
  {http://arxiv.org/abs/hep-th/9904180} {arXiv:hep-th/9904180 [hep-th]}
  \BibitemShut {NoStop}%
\bibitem [{\citenamefont {Huang}\ \emph {et~al.}(2007)\citenamefont {Huang},
  \citenamefont {Hao},\ and\ \citenamefont {Zhuang}}]{Huang:2006fk}%
  \BibitemOpen
  \bibfield  {author} {\bibinfo {author} {\bibfnamefont {X.-G.}\ \bibnamefont
  {Huang}}, \bibinfo {author} {\bibfnamefont {X.-W.}\ \bibnamefont {Hao}}, \
  and\ \bibinfo {author} {\bibfnamefont {P.-F.}\ \bibnamefont {Zhuang}},\
  }\href {\doibase 10.1016/j.astropartphys.2007.09.002} {\bibfield  {journal}
  {\bibinfo  {journal} {Astropart. Phys.}\ }\textbf {\bibinfo {volume} {28}},\
  \bibinfo {pages} {472} (\bibinfo {year} {2007})},\ \Eprint
  {http://arxiv.org/abs/hep-ph/0602186} {arXiv:hep-ph/0602186 [hep-ph]}
  \BibitemShut {NoStop}%
\bibitem [{\citenamefont {Schutzhold}(2002)}]{Schutzhold:2002pr}%
  \BibitemOpen
  \bibfield  {author} {\bibinfo {author} {\bibfnamefont {R.}~\bibnamefont
  {Schutzhold}},\ }\href {\doibase 10.1103/PhysRevLett.89.081302} {\bibfield
  {journal} {\bibinfo  {journal} {Phys. Rev. Lett.}\ }\textbf {\bibinfo
  {volume} {89}},\ \bibinfo {pages} {081302} (\bibinfo {year} {2002})},\
  \Eprint {http://arxiv.org/abs/gr-qc/0204018} {arXiv:gr-qc/0204018 [gr-qc]}
  \BibitemShut {NoStop}%
\bibitem [{\citenamefont {Urban}\ and\ \citenamefont
  {Zhitnitsky}(2010)}]{Urban:2009yg}%
  \BibitemOpen
  \bibfield  {author} {\bibinfo {author} {\bibfnamefont {F.~R.}\ \bibnamefont
  {Urban}}\ and\ \bibinfo {author} {\bibfnamefont {A.~R.}\ \bibnamefont
  {Zhitnitsky}},\ }\href {\doibase 10.1016/j.nuclphysb.2010.04.001} {\bibfield
  {journal} {\bibinfo  {journal} {Nucl. Phys.}\ }\textbf {\bibinfo {volume}
  {B835}},\ \bibinfo {pages} {135} (\bibinfo {year} {2010})},\ \Eprint
  {http://arxiv.org/abs/0909.2684} {arXiv:0909.2684 [astro-ph.CO]} \BibitemShut
  {NoStop}%
\bibitem [{\citenamefont {Flachi}\ and\ \citenamefont
  {Tanaka}(2011)}]{Flachi:2011sx}%
  \BibitemOpen
  \bibfield  {author} {\bibinfo {author} {\bibfnamefont {A.}~\bibnamefont
  {Flachi}}\ and\ \bibinfo {author} {\bibfnamefont {T.}~\bibnamefont
  {Tanaka}},\ }\href {\doibase 10.1103/PhysRevD.84.061503} {\bibfield
  {journal} {\bibinfo  {journal} {Phys. Rev.}\ }\textbf {\bibinfo {volume}
  {D84}},\ \bibinfo {pages} {061503} (\bibinfo {year} {2011})},\ \Eprint
  {http://arxiv.org/abs/1106.3991} {arXiv:1106.3991 [hep-th]} \BibitemShut
  {NoStop}%
\bibitem [{\citenamefont {Flachi}(2013)}]{Flachi:2013iia}%
  \BibitemOpen
  \bibfield  {author} {\bibinfo {author} {\bibfnamefont {A.}~\bibnamefont
  {Flachi}},\ }\href {\doibase 10.1103/PhysRevD.88.041501} {\bibfield
  {journal} {\bibinfo  {journal} {Phys. Rev.}\ }\textbf {\bibinfo {volume}
  {D88}},\ \bibinfo {pages} {041501} (\bibinfo {year} {2013})},\ \Eprint
  {http://arxiv.org/abs/1305.5348} {arXiv:1305.5348 [hep-th]} \BibitemShut
  {NoStop}%
\bibitem [{\citenamefont {Flachi}\ and\ \citenamefont
  {Fukushima}(2014)}]{Flachi:2014jra}%
  \BibitemOpen
  \bibfield  {author} {\bibinfo {author} {\bibfnamefont {A.}~\bibnamefont
  {Flachi}}\ and\ \bibinfo {author} {\bibfnamefont {K.}~\bibnamefont
  {Fukushima}},\ }\href {\doibase 10.1103/PhysRevLett.113.091102} {\bibfield
  {journal} {\bibinfo  {journal} {Phys. Rev. Lett.}\ }\textbf {\bibinfo
  {volume} {113}},\ \bibinfo {pages} {091102} (\bibinfo {year} {2014})},\
  \Eprint {http://arxiv.org/abs/1406.6548} {arXiv:1406.6548 [hep-th]}
  \BibitemShut {NoStop}%
\bibitem [{\citenamefont {Benic}\ and\ \citenamefont
  {Fukushima}(2015)}]{Benic:2015qha}%
  \BibitemOpen
  \bibfield  {author} {\bibinfo {author} {\bibfnamefont {S.}~\bibnamefont
  {Benic}}\ and\ \bibinfo {author} {\bibfnamefont {K.}~\bibnamefont
  {Fukushima}},\ }\href@noop {} {\  (\bibinfo {year} {2015})},\ \Eprint
  {http://arxiv.org/abs/1503.05790} {arXiv:1503.05790 [hep-th]} \BibitemShut
  {NoStop}%
\bibitem [{\citenamefont {Yamamoto}\ and\ \citenamefont
  {Hirono}(2013)}]{Yamamoto:2013zwa}%
  \BibitemOpen
  \bibfield  {author} {\bibinfo {author} {\bibfnamefont {A.}~\bibnamefont
  {Yamamoto}}\ and\ \bibinfo {author} {\bibfnamefont {Y.}~\bibnamefont
  {Hirono}},\ }\href {\doibase 10.1103/PhysRevLett.111.081601} {\bibfield
  {journal} {\bibinfo  {journal} {Phys. Rev. Lett.}\ }\textbf {\bibinfo
  {volume} {111}},\ \bibinfo {pages} {081601} (\bibinfo {year} {2013})},\
  \Eprint {http://arxiv.org/abs/1303.6292} {arXiv:1303.6292 [hep-lat]}
  \BibitemShut {NoStop}%
\bibitem [{\citenamefont {Yamamoto}(2014)}]{Yamamoto:2014vda}%
  \BibitemOpen
  \bibfield  {author} {\bibinfo {author} {\bibfnamefont {A.}~\bibnamefont
  {Yamamoto}},\ }\href {\doibase 10.1103/PhysRevD.90.054510} {\bibfield
  {journal} {\bibinfo  {journal} {Phys. Rev.}\ }\textbf {\bibinfo {volume}
  {D90}},\ \bibinfo {pages} {054510} (\bibinfo {year} {2014})},\ \Eprint
  {http://arxiv.org/abs/1405.6665} {arXiv:1405.6665 [hep-lat]} \BibitemShut
  {NoStop}%
\bibitem [{\citenamefont {Benic}\ and\ \citenamefont
  {Yamamoto}(2016)}]{Benic:2016kdk}%
  \BibitemOpen
  \bibfield  {author} {\bibinfo {author} {\bibfnamefont {S.}~\bibnamefont
  {Benic}}\ and\ \bibinfo {author} {\bibfnamefont {A.}~\bibnamefont
  {Yamamoto}},\ }\href@noop {} {\  (\bibinfo {year} {2016})},\ \Eprint
  {http://arxiv.org/abs/1603.00716} {arXiv:1603.00716 [hep-lat]} \BibitemShut
  {NoStop}%
\bibitem [{\citenamefont {Davies}\ \emph {et~al.}(1996)\citenamefont {Davies},
  \citenamefont {Dray},\ and\ \citenamefont {Manogue}}]{Davies:1996ks}%
  \BibitemOpen
  \bibfield  {author} {\bibinfo {author} {\bibfnamefont {P.~C.~W.}\
  \bibnamefont {Davies}}, \bibinfo {author} {\bibfnamefont {T.}~\bibnamefont
  {Dray}}, \ and\ \bibinfo {author} {\bibfnamefont {C.~A.}\ \bibnamefont
  {Manogue}},\ }\href {\doibase 10.1103/PhysRevD.53.4382} {\bibfield  {journal}
  {\bibinfo  {journal} {Phys. Rev.}\ }\textbf {\bibinfo {volume} {D53}},\
  \bibinfo {pages} {4382} (\bibinfo {year} {1996})},\ \Eprint
  {http://arxiv.org/abs/gr-qc/9601034} {arXiv:gr-qc/9601034 [gr-qc]}
  \BibitemShut {NoStop}%
\bibitem [{\citenamefont {Duffy}\ and\ \citenamefont
  {Ottewill}(2003)}]{Duffy:2002ss}%
  \BibitemOpen
  \bibfield  {author} {\bibinfo {author} {\bibfnamefont {G.}~\bibnamefont
  {Duffy}}\ and\ \bibinfo {author} {\bibfnamefont {A.~C.}\ \bibnamefont
  {Ottewill}},\ }\href {\doibase 10.1103/PhysRevD.67.044002} {\bibfield
  {journal} {\bibinfo  {journal} {Phys. Rev.}\ }\textbf {\bibinfo {volume}
  {D67}},\ \bibinfo {pages} {044002} (\bibinfo {year} {2003})},\ \Eprint
  {http://arxiv.org/abs/hep-th/0211096} {arXiv:hep-th/0211096 [hep-th]}
  \BibitemShut {NoStop}%
\bibitem [{\citenamefont {Nambu}\ and\ \citenamefont
  {Jona-Lasinio}(1961)}]{Nambu:1961tp}%
  \BibitemOpen
  \bibfield  {author} {\bibinfo {author} {\bibfnamefont {Y.}~\bibnamefont
  {Nambu}}\ and\ \bibinfo {author} {\bibfnamefont {G.}~\bibnamefont
  {Jona-Lasinio}},\ }\href {\doibase 10.1103/PhysRev.122.345} {\bibfield
  {journal} {\bibinfo  {journal} {Phys. Rev.}\ }\textbf {\bibinfo {volume}
  {122}},\ \bibinfo {pages} {345} (\bibinfo {year} {1961})}\BibitemShut
  {NoStop}%
\bibitem [{\citenamefont {Schwinger}(1951)}]{Schwinger:1951nm}%
  \BibitemOpen
  \bibfield  {author} {\bibinfo {author} {\bibfnamefont {J.~S.}\ \bibnamefont
  {Schwinger}},\ }\href {\doibase 10.1103/PhysRev.82.664} {\bibfield  {journal}
  {\bibinfo  {journal} {Phys. Rev.}\ }\textbf {\bibinfo {volume} {82}},\
  \bibinfo {pages} {664} (\bibinfo {year} {1951})}\BibitemShut {NoStop}%
\bibitem [{\citenamefont {Gorbar}\ \emph {et~al.}(2011)\citenamefont {Gorbar},
  \citenamefont {Miransky},\ and\ \citenamefont {Shovkovy}}]{Gorbar:2011ya}%
  \BibitemOpen
  \bibfield  {author} {\bibinfo {author} {\bibfnamefont {E.~V.}\ \bibnamefont
  {Gorbar}}, \bibinfo {author} {\bibfnamefont {V.~A.}\ \bibnamefont
  {Miransky}}, \ and\ \bibinfo {author} {\bibfnamefont {I.~A.}\ \bibnamefont
  {Shovkovy}},\ }\href {\doibase 10.1103/PhysRevD.83.085003} {\bibfield
  {journal} {\bibinfo  {journal} {Phys. Rev.}\ }\textbf {\bibinfo {volume}
  {D83}},\ \bibinfo {pages} {085003} (\bibinfo {year} {2011})},\ \Eprint
  {http://arxiv.org/abs/1101.4954} {arXiv:1101.4954 [hep-ph]} \BibitemShut
  {NoStop}%
\bibitem [{\citenamefont {De~Haas}\ and\ \citenamefont {van
  Alphen}(1930)}]{de1930dependence}%
  \BibitemOpen
  \bibfield  {author} {\bibinfo {author} {\bibfnamefont {W.}~\bibnamefont
  {De~Haas}}\ and\ \bibinfo {author} {\bibfnamefont {P.}~\bibnamefont {van
  Alphen}},\ }in\ \href@noop {} {\emph {\bibinfo {booktitle} {Proc. Netherlands
  Roy. Acad. Sci}}},\ Vol.~\bibinfo {volume} {33}\ (\bibinfo {year} {1930})\
  p.\ \bibinfo {pages} {170}\BibitemShut {NoStop}%
\bibitem [{\citenamefont {De~Haas}\ and\ \citenamefont
  {Van~Alphen}(1930)}]{de1930oscillations}%
  \BibitemOpen
  \bibfield  {author} {\bibinfo {author} {\bibfnamefont {W.}~\bibnamefont
  {De~Haas}}\ and\ \bibinfo {author} {\bibfnamefont {P.}~\bibnamefont
  {Van~Alphen}},\ }\href@noop {} {\bibfield  {journal} {\bibinfo  {journal}
  {Leiden Commun}\ }\textbf {\bibinfo {volume} {208}},\ \bibinfo {pages} {212a}
  (\bibinfo {year} {1930})}\BibitemShut {NoStop}%
\bibitem [{\citenamefont {Novoselov}\ \emph {et~al.}(2005)\citenamefont
  {Novoselov}, \citenamefont {Geim}, \citenamefont {Morozov}, \citenamefont
  {Jiang}, \citenamefont {Katsnelson}, \citenamefont {Grigorieva},
  \citenamefont {Dubonos},\ and\ \citenamefont {Firsov}}]{Novoselov:2005kj}%
  \BibitemOpen
  \bibfield  {author} {\bibinfo {author} {\bibfnamefont {K.~S.}\ \bibnamefont
  {Novoselov}}, \bibinfo {author} {\bibfnamefont {A.~K.}\ \bibnamefont {Geim}},
  \bibinfo {author} {\bibfnamefont {S.~V.}\ \bibnamefont {Morozov}}, \bibinfo
  {author} {\bibfnamefont {D.}~\bibnamefont {Jiang}}, \bibinfo {author}
  {\bibfnamefont {M.~I.}\ \bibnamefont {Katsnelson}}, \bibinfo {author}
  {\bibfnamefont {I.~V.}\ \bibnamefont {Grigorieva}}, \bibinfo {author}
  {\bibfnamefont {S.~V.}\ \bibnamefont {Dubonos}}, \ and\ \bibinfo {author}
  {\bibfnamefont {A.~A.}\ \bibnamefont {Firsov}},\ }\href {\doibase
  10.1038/nature04233} {\bibfield  {journal} {\bibinfo  {journal} {Nature}\
  }\textbf {\bibinfo {volume} {438}},\ \bibinfo {pages} {197} (\bibinfo {year}
  {2005})},\ \Eprint {http://arxiv.org/abs/cond-mat/0509330}
  {arXiv:cond-mat/0509330 [cond-mat.mes-hall]} \BibitemShut {NoStop}%
\bibitem [{\citenamefont {Csernai}\ \emph {et~al.}(2011)\citenamefont
  {Csernai}, \citenamefont {Magas}, \citenamefont {Stocker},\ and\
  \citenamefont {Strottman}}]{Csernai:2011gg}%
  \BibitemOpen
  \bibfield  {author} {\bibinfo {author} {\bibfnamefont {L.~P.}\ \bibnamefont
  {Csernai}}, \bibinfo {author} {\bibfnamefont {V.~K.}\ \bibnamefont {Magas}},
  \bibinfo {author} {\bibfnamefont {H.}~\bibnamefont {Stocker}}, \ and\
  \bibinfo {author} {\bibfnamefont {D.~D.}\ \bibnamefont {Strottman}},\ }\href
  {\doibase 10.1103/PhysRevC.84.024914} {\bibfield  {journal} {\bibinfo
  {journal} {Phys. Rev.}\ }\textbf {\bibinfo {volume} {C84}},\ \bibinfo {pages}
  {024914} (\bibinfo {year} {2011})},\ \Eprint {http://arxiv.org/abs/1101.3451}
  {arXiv:1101.3451 [nucl-th]} \BibitemShut {NoStop}%
\bibitem [{\citenamefont {Cook}\ \emph {et~al.}(1994)\citenamefont {Cook},
  \citenamefont {Shapiro},\ and\ \citenamefont {Teukolsky}}]{Cook:1993qr}%
  \BibitemOpen
  \bibfield  {author} {\bibinfo {author} {\bibfnamefont {G.~B.}\ \bibnamefont
  {Cook}}, \bibinfo {author} {\bibfnamefont {S.~L.}\ \bibnamefont {Shapiro}}, \
  and\ \bibinfo {author} {\bibfnamefont {S.~A.}\ \bibnamefont {Teukolsky}},\
  }\href {\doibase 10.1086/173934} {\bibfield  {journal} {\bibinfo  {journal}
  {Astrophys. J.}\ }\textbf {\bibinfo {volume} {424}},\ \bibinfo {pages} {823}
  (\bibinfo {year} {1994})}\BibitemShut {NoStop}%
\bibitem [{\citenamefont {Friedman}\ \emph {et~al.}(1986)\citenamefont
  {Friedman}, \citenamefont {Ipser},\ and\ \citenamefont
  {Parker}}]{friedman1986rapidly}%
  \BibitemOpen
  \bibfield  {author} {\bibinfo {author} {\bibfnamefont {J.}~\bibnamefont
  {Friedman}}, \bibinfo {author} {\bibfnamefont {J.}~\bibnamefont {Ipser}}, \
  and\ \bibinfo {author} {\bibfnamefont {L.}~\bibnamefont {Parker}},\
  }\href@noop {} {\bibfield  {journal} {\bibinfo  {journal} {The Astrophysical
  Journal}\ }\textbf {\bibinfo {volume} {304}},\ \bibinfo {pages} {115}
  (\bibinfo {year} {1986})}\BibitemShut {NoStop}%
\bibitem [{\citenamefont {Glendenning}\ \emph {et~al.}(1997)\citenamefont
  {Glendenning}, \citenamefont {Pei},\ and\ \citenamefont
  {Weber}}]{glendenning1997signal}%
  \BibitemOpen
  \bibfield  {author} {\bibinfo {author} {\bibfnamefont {N.~K.}\ \bibnamefont
  {Glendenning}}, \bibinfo {author} {\bibfnamefont {S.}~\bibnamefont {Pei}}, \
  and\ \bibinfo {author} {\bibfnamefont {F.}~\bibnamefont {Weber}},\
  }\href@noop {} {\bibfield  {journal} {\bibinfo  {journal} {Phys. Rev. Lett.}\
  }\textbf {\bibinfo {volume} {79}},\ \bibinfo {pages} {1603} (\bibinfo {year}
  {1997})}\BibitemShut {NoStop}%
\bibitem [{\citenamefont {Chubarian}\ \emph {et~al.}(1999)\citenamefont
  {Chubarian}, \citenamefont {Grigorian}, \citenamefont {Poghosyan},\ and\
  \citenamefont {Blaschke}}]{chubarian1999deconfinement}%
  \BibitemOpen
  \bibfield  {author} {\bibinfo {author} {\bibfnamefont {E.}~\bibnamefont
  {Chubarian}}, \bibinfo {author} {\bibfnamefont {H.}~\bibnamefont
  {Grigorian}}, \bibinfo {author} {\bibfnamefont {G.}~\bibnamefont
  {Poghosyan}}, \ and\ \bibinfo {author} {\bibfnamefont {D.}~\bibnamefont
  {Blaschke}},\ }\href@noop {} {\  (\bibinfo {year} {1999})},\ \Eprint
  {http://arxiv.org/abs/astro-ph/9903489} {arXiv:astro-ph/9903489 [astro-ph]}
  \BibitemShut {NoStop}%
\bibitem [{\citenamefont {Haensel}\ \emph {et~al.}(2007)\citenamefont
  {Haensel}, \citenamefont {Potekhin},\ and\ \citenamefont
  {Yakovlev}}]{haensel2007neutron}%
  \BibitemOpen
  \bibfield  {author} {\bibinfo {author} {\bibfnamefont {P.}~\bibnamefont
  {Haensel}}, \bibinfo {author} {\bibfnamefont {A.~Y.}\ \bibnamefont
  {Potekhin}}, \ and\ \bibinfo {author} {\bibfnamefont {D.~G.}\ \bibnamefont
  {Yakovlev}},\ }\href@noop {} {\emph {\bibinfo {title} {Neutron stars 1:
  Equation of state and structure}}},\ Vol.\ \bibinfo {volume} {326}\ (\bibinfo
   {publisher} {Springer Science \& Business Media},\ \bibinfo {year}
  {2007})\BibitemShut {NoStop}%
\bibitem [{\citenamefont {Buballa}\ and\ \citenamefont
  {Carignano}(2015)}]{Buballa:2014tba}%
  \BibitemOpen
  \bibfield  {author} {\bibinfo {author} {\bibfnamefont {M.}~\bibnamefont
  {Buballa}}\ and\ \bibinfo {author} {\bibfnamefont {S.}~\bibnamefont
  {Carignano}},\ }\href {\doibase 10.1016/j.ppnp.2014.11.001} {\bibfield
  {journal} {\bibinfo  {journal} {Prog. Part. Nucl. Phys.}\ }\textbf {\bibinfo
  {volume} {81}},\ \bibinfo {pages} {39} (\bibinfo {year} {2015})},\ \Eprint
  {http://arxiv.org/abs/1406.1367} {arXiv:1406.1367 [hep-ph]} \BibitemShut
  {NoStop}%
\bibitem [{\citenamefont {Fukushima}\ and\ \citenamefont
  {A.~Morales}(2013)}]{Fukushima:2013zga}%
  \BibitemOpen
  \bibfield  {author} {\bibinfo {author} {\bibfnamefont {K.}~\bibnamefont
  {Fukushima}}\ and\ \bibinfo {author} {\bibfnamefont {P.}~\bibnamefont
  {A.~Morales}},\ }\href {\doibase 10.1103/PhysRevLett.111.051601} {\bibfield
  {journal} {\bibinfo  {journal} {Phys. Rev. Lett.}\ }\textbf {\bibinfo
  {volume} {111}},\ \bibinfo {pages} {051601} (\bibinfo {year} {2013})},\
  \Eprint {http://arxiv.org/abs/1305.4115} {arXiv:1305.4115 [hep-ph]}
  \BibitemShut {NoStop}%
\bibitem [{\citenamefont {Carignano}\ \emph {et~al.}(2015)\citenamefont
  {Carignano}, \citenamefont {Ferrer}, \citenamefont {de~la Incera},\ and\
  \citenamefont {Paulucci}}]{Carignano:2015kda}%
  \BibitemOpen
  \bibfield  {author} {\bibinfo {author} {\bibfnamefont {S.}~\bibnamefont
  {Carignano}}, \bibinfo {author} {\bibfnamefont {E.~J.}\ \bibnamefont
  {Ferrer}}, \bibinfo {author} {\bibfnamefont {V.}~\bibnamefont {de~la
  Incera}}, \ and\ \bibinfo {author} {\bibfnamefont {L.}~\bibnamefont
  {Paulucci}},\ }\href {\doibase 10.1103/PhysRevD.92.105018} {\bibfield
  {journal} {\bibinfo  {journal} {Phys. Rev.}\ }\textbf {\bibinfo {volume}
  {D92}},\ \bibinfo {pages} {105018} (\bibinfo {year} {2015})},\ \Eprint
  {http://arxiv.org/abs/1505.05094} {arXiv:1505.05094 [nucl-th]} \BibitemShut
  {NoStop}%
\end{thebibliography}%

\end{document}